\documentclass[12pt]{article}
\usepackage{amsfonts,amsmath,amssymb,epsf,framed}
\usepackage{amsmath,braket}
\usepackage{graphicx,hyperref,color,comment}
\usepackage{cite}

\usepackage{subfigure}

\edef\restoreparindent{\parindent=\the\parindent\relax}
\usepackage{parskip}
\restoreparindent

\hypersetup{
    bookmarks=true,         
    unicode=false,          
    pdftoolbar=true,        
    pdfmenubar=true,        
    linktocpage=true,       
    pdffitwindow=false,     
    pdfstartview={FitH},    
    pdfnewwindow=true,      
    colorlinks=true,       
    linkcolor=blue,          
    citecolor=blue,        
    filecolor=blue,      
    urlcolor=blue           
}
\numberwithin{equation}{section}									

\let\a=\alpha \let\b=\beta  \let\d=\delta   \let\g=\gamma \let\h=\eta   \let\m=\mu 
  \let\r=\rho \let\s=\sigma       
 \let\D=\Delta     \let\S=\Sigma    
\def\nn{\nonumber}

\def\pa{\partial}

\begin{document}

\begin{titlepage}
\thispagestyle{empty}

\vspace*{-2cm}
\begin{flushright}
RUP-25-16\\
UT-Komaba/25-5\\
\vspace{2.0cm}
\end{flushright}

\bigskip

\begin{center}
\noindent{{\Large \textbf{Fluid Boundary Conditions from AdS/BCFT}}}\\
\vspace{2cm}

Tomohito Shiga$^a$ \, and \, Kenta Suzuki$^b$
\vspace{1cm}\\

{\it ${}^a$Department of Physics, Rikkyo University, Toshima, Tokyo 171-8501, Japan}, \\[6pt]
{\it ${}^b$Graduate School of Arts and Sciences, University of Tokyo, Komaba, Meguro-ku,\\[2pt] Tokyo 153-8902, Japan}

\vspace{1mm}

\bigskip \bigskip
\vskip 3em
\end{center}

\begin{abstract}
In this paper, we initiate our program to classify conformal fluid boundary conditions, by utilizing the fluid/gravity correspondence in the AdS/BCFT correspondence.
The AdS/BCFT correspondence is a conjectured duality between a quantum gravity in asymptotically AdS spacetime with an end-of-the world brane and a boundary conformal field theory (BCFT).
We show that a choice of boundary condition for the metric on the end-of-the world brane naturally leads to specific boundary conditions for velocity field and temperature field of BCFT in the hydrodynamic limit. 
We analyze the Neumann, Dirichlet and Conformal boundary conditions for the metric on the end-of-the world brane, and discuss their implications for conformal fluid boundary conditions.

\end{abstract}

\end{titlepage}

\newpage

\tableofcontents

\section{Introduction}
\label{sec:introduction}

With the development of superstring theory, the Anti--de Sitter/Conformal Field Theory (AdS/CFT) correspondence has been established as a fundamental framework demonstrating the duality between gravitational theories and gauge theories \cite{Maldacena:1997re,Gubser:1998bc,Witten:1998qj}. In the large $N$ limit, this correspondence suggests that Einstein gravity in a $(d+1)$-dimensional AdS spacetime is equivalent to a $d$-dimensional conformal field theory (CFT) on its asymptotic AdS boundary, enabling the analysis of gauge theories in the strongly coupled regime.

In the hydrodynamic limit (i.e. long-distance and low-frequency limit), this correspondence successfully derived the shear mode and the sound mode of $\mathcal{N}=4$ super-Yang-Mills plasma \cite{Policastro:2002se, Policastro:2002tn}.
A further development in this direction is the derivation of the ratio of shear viscosity $\eta$ to entropy density $s$ in a strongly coupled quark–gluon plasma (QGP). Experimental results at the Relativistic Heavy Ion Collider (RHIC) indicate that the QGP exhibits an extremely low viscosity ratio, close to the conjectured lower bound \cite{Shuryak:2003xe,Shuryak:2004cy,Shuryak:2006se}. In the AdS/CFT framework, one computes the boundary retarded Green’s functions through the GKP–Witten relation between the generating functionals of the gauge and gravity sides:
\begin{align}
  Z_{\mathrm{gauge}} \, = \, Z_{\mathrm{gravity}} \, .
\end{align}
By evaluating these functions, one reproduces the universal result \cite{Kovtun:2004de}
\begin{equation}
  \frac{\eta}{s} \, = \, \frac{1}{4\pi} \, .
\end{equation}
Moreover, by solving the Einstein equations perturbatively in the energy–momentum tensor, the shear viscosity emerges naturally as the coefficient of the dissipative term \cite{Bhattacharyya:2007vjd,Haack:2008cp,Bhattacharyya:2008mz}. These results are in excellent agreement with experimental observations, providing strong evidence for the effectiveness of the AdS/CFT correspondence.

Integrating these insights, the framework that constructs conformal fluid dynamics from small perturbations of the gravitational background is known as the fluid/gravity correspondence
\cite{Ambrosetti:2008mt, Hubeny:2011hd}.
The boosted AdS black brane solution, which corresponds to an ideal conformal fluid at leading order, is given by
\begin{align}
    ds^2 &= -2u_{\mu}dx^{\mu}dr - r^2 f(br) u_{\mu}u_{\nu}dx^{\mu}dx^{\nu} + r^2 P_{\mu\nu}dx^{\mu}dx^{\nu}, \\
    f(r) &= 1 - \frac{1}{r^d} \, .
\end{align}
where the inverse temperature $b$ and fluid velocity $u_{\mu}$ are constants.
Once they are promoted to functions of the CFT coordinates $x^\mu$ and expanded perturbatively in a formal parameter $\epsilon$ as
\begin{align}
    b(\epsilon x^\mu) &= b^{(0)} + \epsilon x^\mu \partial_\mu b + O(\epsilon^2), \\
    u_{\mu}(\epsilon x^\nu) &= u_{\mu}^{(0)} + \epsilon x^\nu \partial_\nu u_{\mu} + O(\epsilon^2) \, ,
\end{align}
the first order in $\epsilon$ corresponds to a viscous conformal fluid.
This procedure yields corrected solutions that include first-order shear terms, allowing holographic analyses of non-equilibrium responses in strongly coupled fluids.

On the other hand, the AdS/BCFT correspondence has been proposed as a gravitational dual to conformal field theory with boundary (BCFT) \cite{Takayanagi:2011zk,Fujita:2011fp}, demonstrating that boundary conditions imposed on the end-of-the-world brane correspond to physical constraints in BCFT.
Although the Neumann boundary condition on the brane (which seems more natural from the junction condition of the brane world holography \cite{Randall:1999ee, Randall:1999vf, Karch:2000ct}) have been predominantly studied in AdS/BCFT \cite{Karch:2000gx, Takayanagi:2011zk,Fujita:2011fp,Nozaki:2012qd,Miyaji:2021ktr,Suzuki:2022xwv,Izumi:2022opi}, there have been recent attempts to impose Dirichlet and conformal boundary conditions \cite{Chu:2017aab,Miao:2017gyt,Miao:2018qkc,Chu:2021mvq}.

We aim to analytically clarify how various boundary conditions for the metric imposed on the end-of-the world brane give rise to the corresponding fluid boundary conditions by combining the fluid/gravity correspondence with AdS/BCFT.
Before we start our investigation, we briefly review the fluid/gravity correspondence and the AdS/BCFT correspondence in section~\ref{sec:reviews}.
Then, we first analyze the Neumann boundary condition on the brane focusing on the zero brane tension case in section~\ref{sec:neumann}.
In section~\ref{sec:other}, we discuss other boundary conditions: the Dirichlet boundary condition and conformal boundary condition.
For the case of finite brane tension, we mainly discuss the Neumann condition in a (2+1)-dimensional BTZ black hole example in section~\ref{sec:brane-tension}.
We conclude this paper in section~\ref{sec:conclusions} with a discussion.
Appendix~\ref{app:second-order-CBC} gives a further treatment of conformal boundary condition case,
and Appendix~\ref{app:extrinsic-curvature} summarizes results of the extrinsic curvature for non-zero brane tension case.

\section{Reviews}
\label{sec:reviews}
In this section, we review the fluid/gravity correspondence in section~\ref{sec:fluid/gravity} and the AdS/BCFT correspondence in section~\ref{sec:ads/bcft}
focusing on minimal amount of contents we need in the rest of the paper.

\subsection{Fluid/Gravity correspondence}
\label{sec:fluid/gravity}

In relativistic fluid dynamics (for example, see \cite{Kovtun:2012rj}), the energy–momentum tensor is generally expressed as
\begin{align}
    T_{\mu\nu}=(\mathcal{E}+\mathcal{P})u_\mu u_\nu + \mathcal{P}\, \eta_{\mu\nu} + \tau_{\mu\nu}^\text{dissipative}(\partial u,\partial^2 u,\cdots,\partial\mathcal{P},\partial^2\mathcal{P},\cdots),
\end{align}
where $\mathcal{E}$ denotes the energy density, $\mathcal{P}$ the pressure, and $u_\mu$ the $d$-dimensional fluid velocity, which satisfies $\h^{\mu\nu}u_\mu u_\nu=-1$. The metric $\eta_{\mu\nu}$ represents the flat spacetime metric. The first two terms describe an ideal fluid, while dissipative effects such as viscosity are incorporated through the derivative expansion term $\tau_{\mu\nu}^\text{dissipative}$. Under the constraints of the second law of thermodynamics (i.e., $\nabla_\mu s^\mu\geq 0$) and the choice of the Landau frame ($u_\mu\tau^{\mu\nu}=0$), the first-order derivative correction takes the form
\begin{align}
    \tau^{\mu \nu} = \Delta^{\mu \rho}\Delta^{\nu \sigma}\left[\eta\left(\partial_{\rho} u_{\sigma}+\partial_{\sigma} u_{\rho}-\frac{2}{d-1}\delta_{\rho\sigma}\partial_{\lambda} u^{\lambda}\right)+\zeta\, \delta_{\rho\sigma}\partial_{\lambda} u^{\lambda}\right],\label{eq:energymomentumdissip1}  
\end{align}
where $\eta$ and $\zeta$ are the shear and bulk viscosities, respectively, which are positive transport coefficients. $\Delta^{\mu\nu}=\h^{\mu\nu}+u^{\mu}u^{\nu}$ is the projector onto the space transverse to the $d$-dimensional fluid velocity $u^\mu$. These coefficients are typically determined through phenomenological or experimental approaches.

Next, consider a $d$-dimensional ideal conformal fluid. In such a theory, the energy density and pressure are related by $\mathcal{E}=(d-1)\mathcal{P}$, and when expressed in terms of the inverse temperature $b$, $\mathcal{P}= \frac{1}{b^d}$. the ideal fluid energy–momentum tensor is given by
\begin{align}
    T^{\mu \nu}=\frac{1}{b^{d}}\left( d\, u^{\mu} u^{\nu} + \eta^{\mu \nu}\right).
\end{align}
Via the AdS/CFT correspondence, the corresponding $(d+1)$-dimensional gravity theory is described by a Lorentz-boosted $\mathrm{AdS}_{d+1}$ black brane spacetime
\cite{Bhattacharyya:2007vjd, Ambrosetti:2008mt, Hubeny:2011hd}:
\begin{align}
    ds^2&=-2u_{\mu}dx^{\mu}dr-r^2f(br)u_{\mu}u_{\nu}dx^{\mu}dx^{\nu}+r^2P_{\mu\nu}dx^{\mu}dx^{\nu},\label{eq:cftperfectfluid}\\
&f(r)=1-\frac{1}{r^d},\ \  P_{\mu\nu}=\eta_{\mu\nu}+u_\mu u_\nu\notag
\end{align}
where the inverse temperature $b$ is defined as the inverse of the horizon radius of the black brane, and the horizon radius is defined by the largest positive root of $f(r)$.
Also $u_\mu$ is the $d$-dimensional velocity parameter introduced via the Lorentz boost on the black brane. 

Here, we are working in the ingoing Eddington–Finkelstein coordinates, which are commonly used in the literature on fluid/gravity correspondence. To emphasize this choice explicitly, we denote the time coordinate among $\{x^\mu\}$ by $x^v$, where corresponds to the ingoing null time coordinate.

When both $b$ and $u_\mu$ are constants, the boosted black brane spacetime (\ref{eq:cftperfectfluid}) is an exact solution of the Einstein equations, holographically dual to conformal ideal fluid.
However, one may promote these parameters to be functions of the boundary coordinates $\{x^\mu\}$:
\begin{align}
    (b,u) \longrightarrow (b(x^\mu),\, u(x^\mu)).
\end{align}
At this stage, the metric in Eq.~\eqref{eq:cftperfectfluid} is no longer an exact solution of the Einstein equations. Nonetheless, if the variations with respect to $x^\mu$ are sufficiently small, the metric can be regarded as an approximate solution. Expanding $b$ and $u$ in a derivative expansion, the Einstein equations
\begin{align}
    E_{MN}[g(b(x^\mu),u(x^\mu))]=\sum_{n=1}^\infty \epsilon^n E_{MN}^{(n)}=0
\end{align}
can be solved order by order. Here, $\epsilon$ is a formal expansion parameter that keeps track of the number of derivatives with respect to the boundary coordinates $ x^\mu $. It is introduced to systematically organize the perturbative expansion in the long-wavelength regime, where variations of $ b(x^\mu) $ and $ u_\mu(x^\mu) $ occur over a length scale much larger than the inverse temperature $ b$. In particular, one writes
\begin{align}
    b(\epsilon x^\mu)&=b^{(0)}+\epsilon\, x^\mu\partial_\mu b+O(\epsilon^2),\\
    u_\mu(\epsilon x^\nu)&=u_\mu^{(0)}+\epsilon\, x^\nu\partial_\nu u_\mu+O(\epsilon^2).
\end{align}
Solving the Einstein equations perturbatively yields the first-order correction to the metric\cite{Bhattacharyya:2007vjd}:
\begin{align}
    ds^2=&-2u_{\mu}dx^{\mu}dr-r^2f(br)u_{\mu}u_{\nu}dx^{\mu}dx^{\nu}+r^2P_{\mu\nu}dx^{\mu}dx^{\nu}\notag\\
       &+\epsilon\Biggl[2r^2bF(br)\sigma_{\mu\nu}+\frac{2}{d-1}ru_\mu u_\nu\, \partial_\lambda u^\lambda -ru^\lambda\partial_\lambda(u_\mu u_\nu)\Biggr]dx^\mu dx^\nu+O(\epsilon^2),\\[1mm]
       \sigma^{\mu\nu}&=P^{\mu \rho} P^{\nu \sigma}\partial_{(\rho} u_{\sigma)}-\frac{1}{d-1}P^{\mu \nu}\partial_{\lambda} u^{\lambda} \, , \qquad
       F(r) \, = \, \int_{r}^{\infty} d x \frac{x^{d-1}-1}{x\left(x^{d}-1\right)}.\notag
\end{align}
By applying the AdS/CFT dictionary once more, the viscous fluid energy–momentum tensor on the CFT side is obtained as
\begin{align}
    T_{\mu \nu}=\frac{1}{b^{d}}\left(d\, u_{\mu} u_{\nu}+\eta_{\mu \nu}\right)-\frac{2}{b^{d-1}} \sigma_{\mu \nu}.
\end{align}
From this expression, one deduces that the shear viscosity of the CFT fluid is given by $\eta\propto \frac{1}{b^{d-1}}$. 

In this framework, the Einstein equations in the bulk are solved order by order in a derivative expansion, systematically constructing the bulk metric. At leading order (zeroth order in derivatives), the bulk metric describes a static black brane, and the conservation of the boundary stress tensor is trivially satisfied. At first order in derivatives, the conservation equation leads to the relativistic Euler equations, governing the ideal fluid dynamics. Extending the expansion to second order, viscous corrections appear in the stress tensor, giving rise to the relativistic Navier–Stokes equations with shear viscosity \cite{Bhattacharyya:2007vjd,Haack:2008cp,Bhattacharyya:2008mz}. Thus, the gravitational description naturally incorporates both ideal and dissipative fluid dynamics, with each successive order in the derivative expansion refining the correspondence between the bulk metric and the dual fluid.

In summary, this framework establishes a direct correspondence between the gravitational description and the CFT fluid dynamics, which is commonly referred to as the fluid/gravity correspondence.

In this paper, we adopt the conventional viewpoint that the treatment of boundary conditions for relativistic fluids can follow the same formal structure as in the non-relativistic case, particularly for scalar and vector fields such as the temperature and fluid velocity~\cite{Landau:FM}. While a fully consistent relativistic formulation of such boundary conditions remains under development, this assumption is commonly employed in various studies and suffices for our purposes.

Specifically, for the temperature $T$, we consider the following representative boundary conditions:
\begin{itemize}
  \item \textbf{Dirichlet condition (fixed temperature):}
  \begin{align}
    T\big|_{\text{boundary}} = T_0,
  \end{align}
  where $T_0$ is a prescribed constant.

  \item \textbf{Neumann condition (zero heat flux):}
  \begin{align}
    n^\mu q_\mu \big|_{\text{boundary}} = 0, \quad \text{with} \quad q^\mu = -\kappa\, \partial^\mu T,
  \end{align}
  where $q^\mu$ is the heat flux vector orthogonal to the fluid velocity, and $n^\mu$ is the normal vector to the boundary.
\end{itemize}

\subsection{AdS/BCFT correspondence}
\label{sec:ads/bcft}

In the conventional AdS/CFT correspondence, the boundary of the AdS space $\S$ ($= \partial \mathrm{AdS}$) is identified with the manifold on which the dual CFT is defined. 
For the Poincar\'{e} coordinates of AdS, $\S$ is a manifold without boundary that extends infinitely.

However, when considering a CFT with a boundary (Boundary CFT or BCFT), the CFT possesses a physical boundary $\partial \S$, and an extension of the AdS/CFT correspondence becomes necessary to incorporate this boundary effect.
To construct an AdS spacetime dual to the BCFT, it is essential to introduce a holographic dual corresponding to $\partial \S$ in addition to $\S$. For this purpose, an end-of-the world brane $Q$ is introduced in the bulk of AdS as the holographic dual to $\partial \S$\cite{Takayanagi:2011zk,Fujita:2011fp}. 
This brane is bulk codimension-1 and arranged so that it coincides with the CFT boundary on the asymptotic boundary of the AdS space (i.e., $\partial \mathrm{AdS}  = \S \cup Q$, and $\partial \S =\partial Q$). 
This spacetime structure and the corresponding duality with the BCFT are referred to as the AdS/BCFT correspondence.

In this setup, the gravitational action is given by the Einstein–Hilbert action supplemented by the Gibbons–Hawking boundary term on the brane $Q$:
\begin{equation}
    I=\frac{1}{16 \pi G_{N}} \int_{\textbf{Bulk}} \sqrt{-g}(R-2 \Lambda)+\frac{1}{8 \pi G_{N}} \int_{\S} \sqrt{-\g} K_\S+\frac{1}{8 \pi G_{N}} \int_{Q} \sqrt{-h} (K_Q-T) \label{eq:AdSBCFTaction}
\end{equation}
Here, the constant $T$ represents the tension of the brane $Q$, and if matter is localized on the brane, it appears as a Lagrangian.
If the Lagrangian of the matter on $Q$ is not constant, the constant $T$ in \eqref{eq:AdSBCFTaction} is replaced by $L_Q$, and one introduces the brane action
\begin{align}
  I_Q=\int\sqrt{-h}L_Q
\end{align}
with the corresponding stress–energy tensor defined by
\begin{align}
  {T^Q}^{\alpha\beta}=\frac{2}{\sqrt{-h}}\frac{\delta I_Q}{\delta h_{\alpha\beta}}.
\end{align}
In the present paper, for simplicity, we proceed by by using the constant tension $T$.

In this paper, uppercase Latin indices $\{M,N,\cdots\}$ denote directions in the entire bulk of the AdS spacetime. Among these, the directions excluding the radial direction $r$ are labeled with Greek indices $\{\mu,\nu,\rho,\sigma,\lambda,\cdots\}$. Furthermore, the tangent directions to the brane are denoted by $\{\alpha,\beta,\gamma,\cdots\}$, and the directions tangent to the brane without the radial direction $r$ are denoted by $\{i,j,k,\cdots\}$. Furthermore, we use $\hat{=}$ to denote conditions which are satisfied only on the brane $Q$.

Taking the variation of the action yields the following boundary term on the brane:
\begin{equation}
 \delta I = \int_Q \delta h^{\alpha\beta}\left(K_{\alpha\beta}- Kh_{\alpha\beta} +  Th_{\alpha\beta} \right) \label{eq:BCFTvariation}
\end{equation}
Here, we assumed that the induced metric at the AdS boundary $\S$ is obeying Dirichlet boundary conditions.
In order to satisfy $\delta I \, \hat{=} \, 0$, the following three types of boundary conditions on the brane $Q$ can be considered.

\begin{center}
  \begin{tabular}[h]{|c|c|c|}\hline
    DBC & NBC & CBC \\ \hline
    $\delta h_{\alpha \beta} \, \hat{=} \, 0$
    & $K_{\alpha\beta} \, \hat{=} \, (K-T)h_{\alpha\beta}$ &   
    $\begin{cases}
      K \, \hat{=} \, \frac{d}{d-1}T,\\[2mm]
      \delta h_{\alpha\beta} \, \hat{=} \, 2\sigma(y)h_{\alpha\beta}
    \end{cases}$ \\[2pt] \hline
  \end{tabular}\vspace*{4pt}
\end{center}
The Dirichlet boundary condition (DBC) fixes the metric on the brane. On the other hand, if $\delta h_{\alpha\beta}\neq 0$, 
one imposes the condition that the term in parentheses in \eqref{eq:BCFTvariation} vanishes; this is the Neumann boundary condition (NBC).
The conformal boundary condition (CBC) fixes both the conformal class of the induced metric, and the trace of the extrinsic curvature.
Therefore, it requires that the conformal structure of the boundary is preserved under variations $\delta h_{\alpha\beta}$. 
We can show explicitly the boundary term vanishes for CDC.
\begin{align}
    \delta I &=-2\sigma  \int_Q h^{\alpha\beta}\left(K_{\alpha\beta}- Kh_{\alpha\beta} +  Th_{\alpha\beta} \right) \notag \\ 
    &= -2\sigma\int_Q \left( K - dK + d  T \right) \notag \\
    &\hspace{4pt} \hat{=} \quad 0 \, .
\end{align}

 It is known that imposing DBC does not guarantee the ellipticity of the equations of motion and leads to difficulties in constraining zero modes, 
 potentially resulting in a breakdown of perturbation theory in quantum gravity. 
 Therefore, as an alternative close to DBC, the conformal boundary condition (CBC), which preserves the conformal structure of the boundary, has been proposed \cite{Anderson_2008,Witten:2018lgb}.
Recent investigations have provided further support for this approach. For example, CBCs have been shown to lead to a well-posed boundary value problem and consistent thermodynamic behavior in flat space gravity \cite{Liu:2024ymn,Anninos:2023epi,Banihashemi:2024yye}.
In Lorentzian signature, the corresponding initial-boundary value problem is more subtle and has been studied from the PDE/geometry viewpoint (including well-posed geometric boundary data and cavity ill-posedness) \cite{Fournodavlos:2021eye,An:2021fcq,An:2025rlw,An:2025gvr,Liu:2025xij}.
In this broader context, related discussions of finite conformal boundaries and observatory setups in (A)dS with CBC can be found in \cite{Anninos:2024wpy,Anninos:2024xhc}.

In studies of AdS/BCFT, the imposition of NBC on the brane has been investigated most extensively \cite{Karch:2000gx, Takayanagi:2011zk,Fujita:2011fp,Nozaki:2012qd,Miyaji:2021ktr,Suzuki:2022xwv,Izumi:2022opi}. 
More recently, research on DBC and CBC has also been carried out \cite{Chu:2017aab ,Chu:2018ntx,Miao:2017gyt,Miao:2018qkc,Chu:2021mvq}, 
but these discussions remain an active area of development. 
In this paper, we examine the consequences of these boundary conditions imposed on the brane within the fluid/gravity correspondence.

\section{Neumann Boundary Condition}
\label{sec:neumann}

First, we consider the Neumann boundary condition (NBC). 
Here, we first review the zero-th order in the sense of the fluid/gravity correspondence \cite{Bhattacharyya:2007vjd,Haack:2008cp,Bhattacharyya:2008mz}, where we study the static (i.e. non-boosted) black hole solution in AdS spacetime.
This is nothing but the ordinary setup studied in the context of the AdS/BCFT correspondence \cite{Takayanagi:2011zk,Fujita:2011fp},
where the bulk Einstein equation and the boundary condition on the brane $Q$ are exactly solved.
In the next subsections, we will solve these equations perturbatively order by order as in the ordinary fluid/gravity correspondence \cite{Bhattacharyya:2007vjd,Haack:2008cp,Bhattacharyya:2008mz}.

In this analysis, we work in a $ (d+1)$-dimensional spacetime and, for simplicity, begin with the case where $T=0$. In this situation, the configuration of the brane depends only on the radial coordinate $ r$ of the AdS spacetime and one spatial coordinate $ w := x^w$.

The metric of the static (i.e. non-boosted) black hole solution is given by
\begin{align}
    ds^2 \, = \, - r^2f(br) dt^2 + \frac{dr^2}{r^2f(br)} + r^2 dx^{a}dx_{a} \, \label{eq:LorentzBH}
\end{align}
and we can show that imposing the NBC with $ T=0$ forces the brane to be located at $ w = \mathrm{Const.}$ \cite{Takayanagi:2011zk,Fujita:2011fp}.
Without loss of generality we set this constant to $ w = 0$, so that the boundary of the CFT is given by the plane on $ w = 0$.
We denote the CFT coordinates by $ \{ x^\mu \} = \{ t, x^a, w \}$ \footnote{This means that here we parametrize the coordinates as $ \{ x^i \} = \{ t, x^{a} \}$.}.
We assume that the position of the brane does not depend on $ t$ or $ x^a$ and can be expressed as $ w = w(r)$. Under this assumption, the normal vector to the brane is given by
\begin{align}
  \left( n^{r}, n^{w}, n^{i}\right) \, = \, \left( -r^4 f(br) w^{\prime}(r), 1,0\right) \cdot \frac{1}{r\sqrt{1+ r^4 f(br) w^{\prime}(r)^{2}}} \, ,
  \label{eq:normEBH}
\end{align}
Using this normal vector, the extrinsic curvature tensor is defined as
\begin{align}
  K_{\a\b} \, = \, h^M_{\ \alpha}h^N_{\ \beta}\nabla_M n_N \, , 
\end{align}
where the projection operator is defined by 
\begin{align}
    h_{MN} \, = \, g_{MN} \, - \, n_M n_N \, . 
\end{align}
Although this expression is written in $ (d+1)$-dimensional coordinates, it can be reduced to $ d$ dimensions due to the condition $ K_{AB}n^B=0$. Using this definition, the components of the extrinsic curvature tensor are given by
\begin{align}
  K_{rr} \, &= \, \frac{-2 r f  w^{\prime\prime} - w^\prime\left(2 f \left(3+{ r^4 f  w^\prime}^2\right)+ br f^\prime \right)}{2 f \left(1+ r^4 f { w^\prime}^2 \right)^{5 / 2}}\\
  K_{tt} \, &= \, \frac{ r^4 f w^\prime \left(2 f + br f^\prime \right)}{2 \sqrt{1+ r^4 f {w^\prime}^2}}\\
  K_{ab} \, &= \, -\frac{ r^4 f w^\prime \d_{ab}}{\sqrt{1+r^4 f {w^\prime}^2}}\\
  K \, &= \, -\frac{r^2\left(2 r f  w^{\prime\prime}+ (2d f + br f') r^4 f {w^\prime}^3 +2((d+2)f+br f') w'\right)}{2\left(1+ r^4 f {w^\prime}^2\right)^{3 / 2}}\label{eq:extCurv}
\end{align}
with all other components vanishing. Here, $ w^\prime$ and $ w^{\prime\prime}$ denote the first and second derivatives of $ w$ with respect to $ r$.

Taking the trace of the boundary condition on the brane,
\begin{equation}
K_{\alpha \beta} \, \hat{=} \, (K-T) h _{\alpha \beta},\label{eq:JunctionCondition}
\end{equation}
yields the relation between $ K$ and $ T$:
\begin{align}
K \, \hat{=} \, \frac{d}{d-1} T.
\end{align}
In particular, when $ T=0$ we must impose the boundary conditions
\begin{align}
\left\{\begin{array}{l}
  K_{\alpha \beta} \, \hat{=} \, 0,\\[2mm]
  K\, \hat{=} \, 0,
\end{array}\right.
\end{align}
which, from the expressions for the extrinsic curvature tensor, imply that
\begin{align}
  w^\prime \ \hat{=} \ 0 \, ,\quad w^{\prime\prime} \ \hat{=} \ 0 \, .
\end{align}
This result indicates that $ w$ must be constant, and without loss of generality, we set
\begin{align}
  w \ \hat{=} \ 0 \, ,
\end{align}
which implies that the brane is arranged perpendicularly to the boundary of the AdS spacetime. 
Consequently, the normal vector to the boundary is simply given by
\begin{align}
  n_M = A \delta^w_M,
\end{align}
where $ A$ is a normalization constant.
Moreover, the projection operator onto the brane, $ h _{MN}$, is defined using this normal vector as
\begin{align}
  h _{MN}=g_{MN}-n_M n_N=g_{MN}-A^2\delta^w_M\delta^w_N \label{eq:projector},
\end{align}
and its $ (\alpha,\beta)$ components serve as the induced metric. Using the normal vector and the projection operator, the extrinsic curvature tensor and its trace are defined by
\begin{align}
  K_{\alpha \beta} \, =& \, h ^M_{\ \alpha}h ^N_{\ \beta}\nabla_M n_N,\\[2mm]
  K \, =& \, h ^{\alpha \beta}K_{\alpha \beta}.
\end{align}
Note that when taking covariant derivatives or contracting indices with the projection operator, one must account for all spacetime components\footnote{When taking the trace, the induced metric is used rather than the projection operator.}. 

The metric for the fluid/gravity correspondence in $ (d+1)$ dimensions up to first order is given by
\begin{align}
  ds^{2}= & -2 u_{\mu} dx^{\mu} dr + r^{2}\left(\eta_{\mu \nu}+(1-f(b r)) u_{\mu} u_{\nu}\right) dx^{\mu} dx^{\nu} \nn\\
  & +\left[2 r^{2} b F(b r) \sigma_{\mu \nu}+\frac{2}{d-1} r u_{\mu} u_{\nu} \partial_{\lambda} u^{\lambda}-r u^{\lambda} \partial_{\lambda}\left(u_{\mu} u_{\nu}\right)\right] dx^{\mu} dx^{\nu}.\\
   F(r) \,& = \, \int_{r}^{\infty} d x \frac{x^{d-1}-1}{x\left(x^{d}-1\right)}.\notag
\end{align}

Since we are solving the boundary condition on the brane $Q$ perturbatively, the brane location is still given by $w=0$ as in the zero-th order case. This justifies that the induced metric on the brane, according to Eq.~\eqref{eq:projector}, is simply $ h _{i j}=g_{i j}$, so that
\begin{align}
h _{\alpha \beta}dx^{\alpha }dx^{\beta} = & -2 u_{i} dx^{i} dr + r^{2}\left(\eta_{i j}+(1-f(b r)) u_{i} u_{j}\right) dx^{i} dx^{j} \notag \\
& +\left[2 r^{2} b F(b r) \sigma_{i j}+\frac{2}{d-1} r u_{i} u_{j} \partial_{\lambda} u^{\lambda}-r u^{\lambda} \partial_{\lambda}\left(u_{i} u_{j}\right)\right] dx^{i} dx^{j}. \label{eq:induce}
\end{align}
It is immediately apparent that the second term represents the dissipative effects of the fluid, as it contains first derivatives of the velocity field. We assume that, within the fluid/gravity correspondence setup, the derivatives of the velocity field and the inverse temperature are small. Consequently, the above metric represents a perturbed boosted black brane (i.e., the first line of Eq.~\eqref{eq:induce} corresponds to the background spacetime when $ u$ and $ b$ are constant).

In general, when perturbations are present the brane may behave dynamically such that $ w^\prime \neq 0$. 
In that case, the brane tension $ T$ becomes finite and the brane configuration becomes nontrivial. For simplicity, we assume that even in the presence of derivative perturbations of the fluid, the brane configuration remains unchanged, so that $ w^\prime = 0$ at all times.
Under this assumption, when $ T=0$, we must impose the boundary conditions
\begin{align}
\left\{\begin{array}{l}
  K_{\alpha \beta} \, \hat{=} \, 0,\\[2mm]
  K\, \hat{=} \, 0,
\end{array}\right.
\end{align}
which require the vanishing of the extrinsic curvature. Note that the extrinsic curvature of the EOW brane can be readily derived from Eq.~\eqref{eq:induce}.

\subsection{Ideal fluid}
\label{sec:ideal}
Now, we consider the zeroth-order contribution of the perturbations (i.e. ignoring derivative terms of $u^\mu$ and $b$, so that only the contribution of the ideal fluid is taken into account). In this case, the induced metric is simply given by
\begin{align}
h _{\alpha \beta}dx^{\alpha }dx^{\beta} =  -2 u_{i} dx^{i} dr + r^{2}\left(\eta_{i j}+(1-f(b r)) u_{i} u_{j}\right) dx^{i} dx^{j} \, .
\end{align}
Then the extrinsic curvature is computed as
\begin{align}
&K^{(0)}_{rr}=0, \\
&K^{(0)}_{r i}=-\frac{u_{w} u_{i}}{\sqrt{1+u_{w}^{2}}}, \\
&K^{(0)}_{ij}=\frac{r^2 u^w\left( \eta_{ij}+u_{i} u_{j}\left(1-f \right)\right)}{ \sqrt{1+u^{w^2}}}-\frac{b r^3 u^wu_{i} u_{j}f^\prime }{2 \sqrt{1+u^{w^2}}}, \\
&K^{(0)}=  \frac{d u^{w}}{\sqrt{1+u_{w}^{2}}}.\label{eq:K0}
\end{align}
In order for NBC to be satisfied, all these components must vanish for any $r$. Consequently, the velocity field must obey the condition
\begin{equation}
u_{w} \, \hat{=} \, u^{w} \, \hat{=} \ 0, \label{eq:normvec}
\end{equation}
which implies that the normal component of the velocity on the brane is zero—identical to the boundary condition for a fixed wall. Moreover, the boundary condition also reflects that the global Lorentz symmetry in the $x^w$ direction is broken due to the configuration of the brane.

\subsection{Viscous fluid}
\label{sec:viscous}

Based on the condition \eqref{eq:normvec} obtained for the perfect fluid case, we next examine the boundary conditions for viscous fluid up to the first order in the perturbation expansion.  
First, the first-order contributions to the extrinsic curvature are computed, yielding the following results.
\begin{align}
K^{(1)}_{rr} \ &\hat{=} \ 0 \, ,
\end{align}
\begin{align}    
K^{(1)}_{ri} \ &\hat{=} \ \frac{ u_{i}\, u^{k} \partial_{k} u_w + \partial_{i} u_w - \partial_w u_{i}}{2r} + b\, F\left(\partial_{i} u_w + \partial_w u_{i}\right)\notag\\
&\quad -\frac{1}{2}b\left( u_{i}\, u^{k} \partial_{k} u_w + 2\, \partial_{i} u_w\right)\left(2F + r\, \partial_r F\right)\notag\\
&\quad + b\, F\, u^{k} u_{i}\left(\partial_{k} u_w +  [(1-f(br))-\frac{rb f^\prime}{2}]\, \partial_w u_{k}\right),
\end{align}
\begin{align}
K^{(1)}_{ij} \ &\hat{=} \ -r\left(\eta_{ij} +[(1-f(br))-\frac{rb f^\prime}{2}]\, u_{i} u_{j}\right) u^{k} \partial_{k} u_w - \frac{d\, r\, u_{i} u_{j}\, \partial_w b}{2b}(1-f(br))\notag\\
&\quad +\frac{r\, u_{j}\left(\partial_w u_{i} - \partial_{i} u_w\right)}{2}(1-f(br)) + \frac{r\, u_{i}\left(\partial_w u_{j} - \partial_{j} u_w\right)}{2}(1-f(br)),
\end{align}
\begin{align}
K^{(1)} \ &\hat{=} \ -\frac{1}{r}\left(-2 + d + 2b\,F\,r + b\,r^2\,\partial_r F\right) u^{k}\partial_{k} u_w \nn\\ 
&\quad- \frac{1}{r}\left(1 - b\,F\,r\left(2f+rb f^\prime \right)\right) u^{k}\partial_w u_{k} \, , 
\end{align}
where we have used $u^{w} \, \hat{=} \, 0$.
In order for these first-order contributions to vanish, constraints must be imposed on the velocity field $u^\mu$ and the inverse temperature $b$.  
First, since $\h^{\mu\nu}u_\mu u_\nu=-1$ implies $u^{\mu}\partial_\nu u_{\mu} = u_{\mu}\partial_\nu u^{\mu} = 0$ and we assume that $u_w \, \hat{=} \, 0$ at every point on the brane, it follows that $u^{k}\partial_w u_{k}$ vanishes trivially.
\begin{align}
    &u^\mu\partial_\nu u_\mu =u_\mu\partial_\nu u^\mu=0\notag\\
    &u^i\partial_\nu u_i \, \hat{=} \, u_i\partial_\nu u^i \, \hat{=} \, 0 \quad(\because u^w \, \hat{=} \, u_w \, \hat{=} \, 0)
\end{align}
Furthermore, the condition $u^{w} \, \hat{=} \, 0$ also implies $\partial_{i} u_w \ \hat{=} \, 0$.  
Taking these facts into account, in addition to $K^{(1)}_{rr}$, the quantity $K^{(1)}$ also vanishes trivially. Therefore, the conditions to be examined become
\begin{align}
    K^{(1)}_{ri}:&\quad \left(-\frac{1}{2r} + b\,F\right) \partial_w u_{i} \, \hat{=} \, 0, \label{eq:K1rmu} \\
    K^{(1)}_{ij}:&\quad -\frac{d\,r\, u_{i} u_{j}\,\partial_w b}{b} + r\, u_{j}\,\partial_w u_{i} + r\, u_{i}\,\partial_w u_{j} \, \hat{=} \, 0, \label{eq:K1munu}
\end{align}
which yield the boundary conditions
\begin{align}
    \partial_w u_{i} \, \hat{=} \, 0 \, , \qquad \partial_w b \, \hat{=} \, 0 \, .
\end{align}
Summarizing the boundary conditions obtained so far, we have
\begin{gather}
u_w \ \hat{=} \ 0 \, , \label{eq:uw1}\\[2mm]
\partial_w b \ \hat{=} \ 0 \, , \label{eq:b1}\\[2mm]
\partial_w u_{i} \ \hat{=} \ 0 \, . \label{eq:umu1}
\end{gather}
Condition \eqref{eq:uw1} corresponds, as mentioned earlier, to the boundary condition for a fluid on a fixed wall.  
Condition \eqref{eq:b1} implies that the temperature and pressure gradients in the direction normal to the brane vanish, which is also interpreted as a fluid boundary condition.  
On the other hand, condition \eqref{eq:umu1} states that the gradient of the tangential velocity field in the direction perpendicular to the boundary is zero. This differs from the no-slip condition typically imposed on a viscous fluid at a fixed wall (where the spatial components satisfy $u_{i}=\delta^{v}_{\ i}$) \cite{levlandaufluid}.  
These results indicate that the Neumann boundary condition possesses properties distinct from those of a simple fixed-wall condition.

\section{Other Boundary Conditions}
\label{sec:other}
In this section, we discuss implications of the other two boundary conditions.

\subsection{Dirichlet boundary condition}
\label{sec:dirichlet}
We now discuss the Dirichlet boundary condition (DBC).  
In this case, the boundary condition imposed on the brane is simply  
\begin{align}
    \delta h_{\alpha\beta} \ \hat{=} \ 0 \, .
\end{align}  
Here, as in the NBC case, we consider applying this condition to a brane that is perpendicular to the boundary of the AdS spacetime with $ T=0 $.  
In the fluid/gravity correspondence, the spacetime perturbations are induced by the coordinate dependence of the fluid velocity $u^\mu$ and the inverse temperature $ b $, that is, by the terms appearing in the Taylor expansion about a given spacetime point.  
Under the Dirichlet boundary condition, these contributions vanish entirely, so that $ u^\mu $ and $ b $ take constant values on the brane. Thus, we have  
\begin{align}
    u^i \ \hat{=} \ \textrm{Const}. \qquad b \ \hat{=} \ \textrm{Const}.
\end{align}  
This means that the DBC freezes gravity on the brane.
Moreover, since the global Lorentz symmetry is broken by the brane, the velocity in the direction normal to the brane is, regardless of the boundary condition, set to zero:  
\begin{align}
    u^w \ \hat{=} \ 0 \, .
\end{align}  
 Consequently, when expressing the velocity in the local rest frame, we adopt the convention $ u^v=1 $ with all other components vanishing:  
\begin{align}
    u^\mu \ \hat{=} \ \delta^{\mu}_{v} \, .
\end{align}  
This corresponds to the no-slip boundary condition for a viscous fluid.

We emphasize that here we are studying a set-up of ideal fluid, but we obtained the no-slip boundary condition for a viscous fluid.
This result seems a little bit weird from the phenomenological argument of fluid mechanics, which states that for ideal fluid without viscosity, the parallel velocity $u^i$ cannot have any boundary condition.
This might be related to the non-ellipticity of the Dirichlet boundary condition \cite{Witten:2018lgb}, but we do not investigate this aspect in detail here.

\subsection{Conformal boundary condition}
\label{sec:conformal}

Next, we discuss the conformal boundary condition (CBC).  
The boundary conditions imposed on the brane are given as follows:
\begin{align}
  \begin{cases}
K \, \hat{=}\ \frac{d}{d-1} T \, , &  \\[4pt]
\delta h _{\alpha \beta} \, \hat{=} \ 2 \sigma(y) h _{\alpha \beta} \, .
\label{eq:CDC}
\end{cases}
\end{align}
Here, $\sigma(y)$ is an arbitrary conformal factor, which is a function of the coordinates of the brane $y^\a$.

Let us first study the static black hole background (\ref{eq:LorentzBH}).
In this case, there is no perturbation for the metric, so that $\delta h _{\alpha \beta}=0$ and $\s(y)=0$.
More generally, even for CBCs one can allow a nontrivial conformal factor that depends on intrinsic coordinates on the brane
(including time when the hypersurface is timelike as discussed in \cite{Galante:2025tnt}),
and we present such a discussion in Appendix \ref{app:non-trivial-CBC} for our setup.
Therefore, our solution $\s(y)=0$ should be viewed as a choice rather than an automatic consequence.
For the boundary condition imposed on the trace of the extrinsic curvature, as in the NBC, we study the case with $T=0$, which implies $K \, \hat{=} \, 0$.
We note that, in contrast to the NBC, we do not require the each component of the extrinsic curvature to vanish, but only require it is traceless.
By using the expression in (\ref{eq:extCurv}), in order to determine the brane location $w=w(r)$, we now need to solve the non-linear second order differential equation 
\begin{align}
    \left(2 r f  w^{\prime\prime}+ (2d f + br f') r^4 f {w^\prime}^3 +2((d+2)f+br f') w'\right) \, = \, 0 \, ,
\end{align}
for $w(r)$.
In the NBC case, the requirement of vanishing $K_{tt}$ and $K_{ab}$ led to $w' \, \hat{=} \, 0$,
but this is not the case in the current CBC case, and we need to directly solve the above differential equation.
We are not sure that $w(r)=$ constant is the {\it unique} solution for the above differential equation, but it is certainly a solution.
In order to simplify the following discussion, in this paper we only consider this solution (i.e. the brane location is given by $w(r)= 0$.)  

Thus, starting from the un-boosted AdS Schwarzschild metric \eqref{eq:LorentzBH}, we impose CBC with $T=0$ and adopt the solution
 \begin{align} 
 w \ \hat{=} \ 0 \, . 
\end{align}
Under this choice, the normal vector to the boundary remains simply
 \begin{align} n_M = A \delta^w_M, 
\end{align} 
where $A$ is a normalization constant, as previously defined.

In AdS/BCFT, the analysis of CBC imposed on the EOW brane is still an underdeveloped area. In particular, there is no established general method for specifying the perturbation $\delta h $, i.e., for determining how it is provided to a specific spacetime. Therefore, in this work we consider two cases for the perturbation:
\begin{itemize}
  \item The metric variation induced by diffeomorphism on the brane, expressed as the Lie derivative $\mathcal{L}_{\xi} h =2\sigma h $.
  \item  The metric variation induced by the fluctuation of the fluid parameters $u^\mu$ and $b$.
\end{itemize}
Although both types of perturbations are comprehensive, a fully systematic treatment of either setup beyond the leading order becomes considerably more involved. 
For clarity, in the main text we focus on the perturbation induced by a diffeomorphism on the brane and analyze its leading-order consequence in the derivative expansion. 
The complementary setup in which the perturbation is induced directly by fluctuations of the fluid data $u^\mu$ and $b$ is presented in Appendix~\ref{app:second-order-CBC}.

In the ideal fluid case, the zeroth-order extrinsic curvature of the brane takes the same form as in \eqref{eq:K0}. 
Using the first equation of the conformal boundary condition (\ref{eq:CDC}), we immediately deduce the condition $u^{w}\hat{=} \ 0$.
The remainder of this section is thus devoted to exploring the implications of the second boundary condition in the ideal fluid scenario.

\subsubsection{Diffeomorphism on the brane}  

Assume that the metric perturbation is provided by a coordinate transformation on the brane, $x\rightarrow x-\xi$.  
In this case, the perturbation is given by the Lie derivative of the metric with respect to the vector $\xi$. Applying the conformal boundary condition, we have
\begin{align}
\delta h _{\alpha \beta}=\mathcal{L}_{\xi}h _{\alpha \beta}=\xi^{ \gamma }\partial_{ \gamma }h _{\alpha \beta}+h _{\alpha  \gamma }\partial_{\beta}\xi^{ \gamma } +h _{ \gamma \beta}\partial_{\alpha }\xi^{ \gamma } =2\sigma\,h _{\alpha \beta}.\label{eq:CKE}
\end{align}
This is known as the conformal Killing equation.

In the derivative expansion of the fluid/gravity correspondence, the induced metric can be organized
by the order of derivative data. In the present analysis we focus on the leading-order constraint,
for which it is sufficient to keep only the zeroth-order induced metric $g_q$:
\begin{align}
h(r,u,b,\partial u,\partial b)_{\alpha\beta}
&=g_q(r,u(x),b(x))_{\alpha\beta}+\mathcal{O}(\epsilon),
\label{eq:h_to_gq_leading}\\
g_q(r,u,b)_{ir}&=-u_i,
\qquad
g_q(r,u,b)_{ij}=r^{2}\Bigl(\eta_{ij}+\bigl(1-f(br)\bigr)u_i u_j\Bigr).
\label{eq:gq_def_main}
\end{align}
More precisely, we use the counting
\begin{align}
\partial_r=\mathcal{O}(\epsilon^0),\qquad
\partial_i u,\ \partial_i b=\mathcal{O}(\epsilon),\qquad
\xi^r,\ \xi^i=\mathcal{O}(\epsilon),
\label{eq:counting_main}
\end{align}
under which derivative (viscous) structures contribute only beyond the leading order.
We present the detailed decomposition including $\mathcal{O}(\epsilon)$ terms and the separation of
orders in Appendix~\ref{app:CKEcounting}.

With \eqref{eq:h_to_gq_leading}, the conformal Killing equation \eqref{eq:CKE} at
$\mathcal{O}(\epsilon)$ reduces to
\begin{align}
\mathcal{O}(\epsilon):\quad
\xi^{r}\partial_{r}{g_{q}}_{\alpha\beta}
+{g_{q}}_{\alpha\gamma}\partial_{\beta}\xi^{\gamma}
+{g_{q}}_{\gamma\beta}\partial_{\alpha}\xi^{\gamma}
=2\sigma\,{g_{q}}_{\alpha\beta}.
\label{eq:CKE_Oe_main}
\end{align}
We now proceed to analyze Eq.~\eqref{eq:CKE_Oe_main}.
In particular, we begin by examining its components, starting with the $(r,r)$ component.
\begin{align}
&{g_{q}}_{r \gamma }\partial_{r}\xi^{ \gamma } +{g_{q}}_{ \gamma r}\partial_{r}\xi^{ \gamma }=0,\notag\\
&{g_{q}}_{r k}\partial_{r}\xi^{k} +{g_{q}}_{kr}\partial_{r}\xi^{k}
=2{g_{q}}_{r k}\partial_{r}\xi^{k}=0,\notag\\
&u_{k}\partial_{r}\xi^{k}=0.
\end{align}
Next, consider the $(i,r)$ component.
\begin{align}
&\xi^{r}\partial_{r}{g_{q} }_{ir}+{g_{q}}_{i \gamma }\partial_{r}\xi^{ \gamma } +{g_{q}}_{ \gamma r}\partial_{i}\xi^{ \gamma }=2\sigma\, {g_{q}}_{ir},\notag\\
&{g_{q}}_{ir}\partial_{r}\xi^{r} +{g_{q}}_{ik}\partial_{r}\xi^{k} +{g_{q}}_{kr}\partial_{i}\xi^{k}=2\sigma\, {g_{q}}_{ir},\notag\\
&u_{i}\partial_{r}\xi^{r} 
+r^2\left(\eta_{ik }+  \frac{u_{i}u_{j}}{b^{d}r^{d}}  \right)\partial_{r}\xi^{k}
+u_{k}\partial_{i}\xi^{k}=-2\sigma\, u_{i},\notag\\
&u_{i}\partial_{r}\xi^{r} + r^2\partial_{r}\xi_{i}+u_{k}\partial_{i}\xi^{k}
=-2\sigma\, u_{i}\quad (\because\ u_{k}\partial_{r}\xi^{k}=0).
\end{align}
Finally, consider the $(i,j)$ component.
\begin{align}
&\xi^{r}\partial_{r}{g_{q} }_{ij}+{g_{q}}_{i \gamma }\partial_{j}\xi^{ \gamma } +{g_{q}}_{ \gamma j}\partial_{i}\xi^{ \gamma }=2\sigma\, {g_{q}}_{ij},\notag\\
&2r\xi^{r}\left(\eta_{ij}+ (1-f(br))u_{i}u_{j}\right)-dr\xi^{r}(1-f(br))u_{i}u_{j}+u_{i}\partial_{j}\xi^{r}  \notag\\
&+r^{2}\left(\eta_{ik}+(1-f(br))u_{i}u_{k}\right)\partial_{j}\xi^{k}+r^{2}\left(\eta_{kj}+(1-f(br))u_{k}u_{j}\right)\partial_{i}\xi^{k }+u_{j}\partial_{i}\xi^{r}\notag\\
&=2\sigma\, r^{2}\left(\eta_{ij}+ (1-f(br))u_{i}u_{j}\right).
\end{align}
Comparing the terms on both sides order by order in $r$, we obtain
\begin{align}
&O(r^{2}):\quad 2r\xi^{r}\eta_{ij}+r^{2}(\partial_{i}\xi_{j} +\partial_{j}\xi_{i})=2\sigma\, r^{2}\eta_{ij}, \\
&O(r^{(2-d)}):\quad (2-d)r\xi^{r}(1-f(br))u_{i}u_{j}=2\sigma\, r^{2}(1-f(br))u_{i}u_{j}, \\
&O(r^{(0)}):\quad u_{i}\partial_{j}\xi^{r} +u_{j}\partial_{i}\xi^{r}=0.
\end{align}
From the $O(r^{(2-d)})$ term, i.e., from the relation $(2-d)r\xi^{r}=2\sigma\, r^{2}$, it follows that for $d=2$ we must have $\sigma=0$, meaning that conformal transformations are not permitted. Hence, we consider the case $d\neq2$. In this case, $\xi^r$ takes the form
\begin{align}
\xi^{r}=\frac{2\sigma}{2-d}r.
\end{align}
This leads, first, to the following relations for the $O(r^{(2)})$ and $O(r^{(0)})$ terms:
\begin{align}
O(r^{(2)}):&\quad \frac{4\sigma}{2-d}\eta_{ij}+\partial_{i}\xi_{j} +\partial_{j}\xi_{i}=2\sigma\, \eta_{ij}, \\[2mm]
&\quad \partial_{i}\xi_{j} +\partial_{j}\xi_{i} = -\frac{2d}{2-d}\sigma\, \eta_{ij},\\[2mm]
O(r^{(0)}):&\quad u_{i}\partial_{j}\xi^{r} +u_{j}\partial_{i}\xi^{r}=0.
\end{align}
Taking the trace of the $O(r^{(2)})$ equation with $\eta^{ij}$ yields
\begin{align}
& \partial_{k}\xi^{k}=  \frac{d}{d-2}\sigma (d-1)= d\frac{d-1}{d-2}\sigma.
\end{align}
Similarly, contracting the $O(r^{(2)})$ equation with $u^{i}$ gives
\begin{align}
u^{i}\partial_{j}\xi_{i} = -\frac{2d}{2-d}\sigma\, u_{j}-u^{i}\partial_{i}\xi_{j}.
\end{align}
Taking the trace of the $O(r^{(0)})$ equation with $\eta^{ij}$ yields
\begin{align}
u^{k}\partial_{k}\sigma=0.
\end{align}
Furthermore, by contracting the $(i,r)$ condition with $u^{i}$, we obtain
\begin{align}
&-\partial_{r}\xi^{r} + r^{2}u^{i}\partial_{r}\xi_{i}+u_{k}u^{i}\partial_{i}\xi^{k}
= 2\sigma, \notag\\
&-\left( \frac{2\sigma}{2-d} + \frac{2r}{2-d}\partial_{r}\sigma  \right) +u_{k}u^{i}\partial_{i}\xi^{k}
=  2\sigma \quad (\because\ u^{i}\partial_{r}\xi_{i}=u_{i}\partial_{r}\xi^{i}=0),\notag \\
&u_{k}u^{i}\partial_{i}\xi^{k}
= 2\sigma\left( 1 + \frac{1}{2-d} \right) + \frac{2r}{2-d}\partial_{r}\sigma,\notag \\
&u_{k}u^{i}\partial_{i}\xi^{k}
=u^{k}u^{i}\partial_{i}\xi_{k}
= 2\sigma\, \frac{3-d}{2-d}  + \frac{2r}{2-d}\partial_{r}\sigma.
\end{align}
Next, multiplying the $O(r^{(2)})$ equation
\begin{align}
&\partial_{i}\xi_{j} +\partial_{j}\xi_{i} = -\frac{2d}{2-d}\sigma\, \eta_{ij},
\end{align}
by $u^{i}u^{j}$ yields
\begin{align}
u^{i}u^{j}\partial_{i}\xi_{j}=  \frac{d}{2-d}\sigma.
\end{align}
Thus,
\begin{align}
u^{i}u^{j}\partial_{i}\xi_{j}=  2\sigma\, \frac{3-d}{2-d}  + \frac{2r}{2-d}\partial_{r}\sigma =  \frac{d}{2-d}\sigma.
\end{align}
Hence,
\begin{align}
\frac{2r}{2-d}\partial_{r}\sigma= \left( \frac{d}{2-d}- 2\frac{3-d}{2-d} \right)\sigma= \left( \frac{3d-6}{2-d} \right)\sigma= -3\sigma,
\end{align}
or equivalently,
\begin{align}
\partial_{r}\sigma = -3\frac{2-d}{2r}\sigma.
\end{align}
This implies that the $r$-derivative of $\xi^{r}$ is given by
\begin{align}
\partial_{r}\xi^{r}=\frac{2\sigma}{2-d} + \frac{2r}{2-d}\partial_{r}\sigma=\frac{2\sigma}{2-d}-3\sigma.
\end{align}
Moreover, the $r$-dependence of $\sigma$ is found to be
\begin{align}
\sigma =A(x^\mu)r^{-3(1- \frac{d}{2})}.
\end{align}
Applying these transformation rules to the $(i,r)$ components, we obtain
\begin{align}
&u_{i}\partial_{r}\xi^{r} + r^2\partial_{r}\xi_{i}+u_{k}\partial_{i}\xi^{k}
=2\sigma\, u_{i},\notag \\
&r^2\partial_{r}\xi_{i}=\left(2\sigma -\partial_{r}\xi^{r} \right)u_{i}-u_{k}\partial_{i}\xi^{k}.
\end{align}
Since $\partial_{r}\xi^{r}=\frac{2\sigma}{2-d}  -3\sigma$ and
\begin{align}
u^{k}\partial_{i}\xi_{k} = u_{k}\partial_{i}\xi^{k} = -\frac{2d}{2-d}\sigma\, u_{i}-u^{k}\partial_{k}\xi_{i},
\end{align}
it follows that
\begin{align}
r^{2}\partial_{r}\xi_{i}&=\left(5- \frac{2}{2-d} \right)\sigma\, u_{i}+ \frac{2d}{2-d}\sigma\, u_{i}+  u^{k}\partial_{k}\xi_{i} \\
&=\frac{8-3d}{2-d}\sigma\, u_{i} + u^{k}\partial_{k}\xi_{i}.
\end{align}
Differentiating both sides with respect to $\partial^{i}$ gives
\begin{align}
r^{2}\partial_{r}\partial^{i}\xi_{i}&= \frac{8-3d}{2-d}\, u_{i}\partial^{i} \sigma  + u^{k}\partial_{k}\partial^{i}\xi_{i}.
\end{align}
Since $\partial_{i}\xi^{i}=\partial^{i}\xi_{i}=d\, \frac{d-1}{d-2}\sigma$ and $u^{i}\partial_{i}\sigma=u_{i}\partial^{i}\sigma=0$, it follows that
\begin{align}
&r^{2}\, d\frac{d-1}{d-2}\partial_{r}\sigma=d\, \frac{d-1}{d-2}\, u^{k}\partial_{k}\sigma, \notag\\
&r^{2}\partial_{r}\sigma=u^{k}\partial_{k}\sigma=0,\notag \\
&\frac{r}{2}(2-d)\sigma=0.
\end{align}
Since $d\geq3$, the only solution is $\sigma=0$.  
Therefore, in the spacetime under consideration, conformal transformations induced by coordinate transformations are not allowed in any dimension, and thus this perturbation does not satisfy the conformal boundary condition.

\section{Non-zero Brane Tension}
\label{sec:brane-tension}
Finally, we attempt an analysis for the case where the brane has nonzero tension $T$. In this section, we only study the Nuemman boundary condition.
First, we start from the static black hole background \eqref{eq:LorentzBH} with assuming brane profile $w=w(r)$.
For the NBC
\begin{align}
  K_{\alpha\beta} \ &\hat{=} \ (K-T)\gamma_{\alpha\beta} \, ,\\
  K \ &\hat{=} \ \frac{d}{d-1}T \, ,
\end{align}
from $(rr)$ component, we find
\begin{align}
  w^\prime \, = \, -\frac{2T}{r^2 \sqrt{4f\left((-1+d)^2f-T^2\right)+4(-1+d)rb\,f\,f^\prime+r^2b^2{f^\prime}^2}} \, .
\end{align}
From $(ra)$ component, we find
\begin{align}
  w^\prime \, = \, -\frac{2T}{r^2\sqrt{4f\left((-1+d)^2f-T^2\right)+4(-1+d)^2rb\,f\,f^\prime+(-1+d)^2r^2b^2{f^\prime}^2}} \, ,
\end{align}
and from $(aa)$ component, we find
\begin{align}
  w^\prime \, = \, -\frac{T}{\sqrt{r^4f\left((-1+d)^2f-T^2\right)}} \, .
\end{align}
When $T\neq0$, we can see that for $d>2$, the brane configuration is not uniquely determined. In the case $d=2$, however, we obtain the well-known BTZ spacetime result \cite{Takayanagi:2011zk,Fujita:2011fp}:
\begin{align}
  w^\prime=-\frac{T}{r^2\sqrt{1-fT^2}}.  \label{eq:BTZwprime}
\end{align}
In this case, the brane configuration $w$ is given by
\begin{align}
  w = b\, \operatorname{Arctanh}\left[\frac{br}{T}\sqrt{1-fT^2}\right].
\label{eq:BTZw}
\end{align}
Thus, to avoid the issue of an undetermined brane configuration in general dimensions and for the sake of simplicity, we perform our analysis in the BTZ black hole spacetime in the rest of this section.
The result for higher dimension indicates that the Neumann boundary condition with $T \ne 0$ is inconsistent with the current simple set-up of AdS/BCFT correspondence and the static black hole solution \eqref{eq:LorentzBH}. One possibility to make the Nuemman boundary condition with $T \ne 0$ in higher dimension consistent would be to introduce a spacetime dependence of the brane tension $T(x)$, for example by inducing a brane-localized field, like discussed in \cite{Kanda:2023zse}.

Now we consider the $2+1$ dimensional spacetime. As noted above, even with a constant brane tension, the brane position can be uniquely fixed when $T\neq0$. Therefore, following the same method as before, we consider the fluid/gravity correspondence metric in $2+1$ dimensions
\begin{align}
  d s^{2}=&-2 u_{\mu} d x^{\mu} d r -r^{2}\left(1-\frac{1}{r^{2} b^{2}}\right)u^{\lambda} u_{\mu} u_{\nu} d x^{\mu} d x^{\nu}+r^{2} P_{\mu \nu} d x^{\mu} d x^{\nu}\notag\\
  &-\frac{2}{r} \partial_{\lambda} u^{\lambda} u_{\mu} u_{\nu} d x^{\mu} d x^{\nu}-r u^{\lambda} \partial_{\lambda}\left(u_{\mu} u_{\nu}\right) d x^{\mu} d x^{\nu}\label{eq:fluidgravityOrder1d=3}
\end{align}
and discuss how the fluid boundary conditions emerge when the NBC is applied.

A key difference from higher-dimensional cases is the absence of a term involving $ F(br) $. 
In the general $ (d+1) $-dimensional formulation of the fluid/gravity correspondence,
 $ F(br) $ originates from the corrections to the metric components that transform as the symmetric traceless part of the SO($d-1$) representation of the spatial rotational group. 
 These components encode the response of the dual fluid to shear deformations and are responsible for viscous effects in the boundary hydrodynamics.

However, in $ 2+1 $ dimensions, the relevant symmetry group is only SO(1), which consists of a single reflection symmetry and does not admit any non-trivial traceless symmetric tensors. 
This implies that no independent off-diagonal components exist in the metric perturbation that could contribute to shear viscosity effects. As a result, the term $ F(br) $,
 which is present in higher-dimensional cases to capture these contributions, does not appear in the $ d=2 $ case. Instead, the leading-order metric remains structurally identical to that of an ideal fluid,
  and higher-order transport corrections associated with viscosity do not emerge at this level.

First, we examine the boundary conditions for the perfect fluid case by neglecting the derivative contributions of $u^\mu$.
We compute the components of the extrinsic curvature tensor under the same setup in which the brane embedding is fixed by Eq.~\eqref{eq:BTZwprime}.
The explicit results are summarized in Appendix~\ref{app:extrinsic-curvature}.
General components of the junction condition~\eqref{eq:JunctionCondition} are extremely involved and generally not analytically solvable.
However, due to the vanishing of the $(v,w)$ component of the extrinsic curvature, the corresponding condition can be written in a simplified form as:
\begin{align}
r^2T(1-f)u_v u_w \ \hat{=} \ 0 \, ,
\end{align}
where, we have used the trace junction condition $K = 2T$. This equation is satisfied only when the velocity field satisfies the same boundary condition as in the tensionless case:
\begin{equation}
  u_w \ \hat{=} \ 0 \, .
\end{equation}
Under this condition, it is straightforward to verify that the remaining components of the junction condition are also satisfied.

Finally, we provide some remarks on the contributions from the first derivatives of $u$ and $b$. As in the previous discussion, assuming that the brane embedding is uniquely determined by Eq.~\eqref{eq:BTZwprime}, one cannot impose additional boundary conditions on the first derivatives of $u$ and $b$. These derivatives are treated as perturbative corrections and, in the presence of a brane with finite tension, their physical effects must be properly taken into account.

In the preceding analysis, we considered a static embedding in which the brane position $w$ depends only on the radial coordinate, i.e., $w = w(r)$. However, for more realistic configurations, one must consider time-dependent profiles of the form $w = w(r, v)$. A detailed analysis of such time-dependent embeddings is left for future work.

\section{Conclusions and Discussions}
\label{sec:conclusions}
In this paper, we examined conformal fluid boundary conditions by combining the fluid/gravity correspondence and the AdS/BCFT correspondence.
We have studied the consequences of the Neumann, Dirichlet, and Conformal boundary conditions for the metric imposed on the end-of-the-world brane.
We first studied zero brane tension case: $T=0$.
For ideal fluid, the Neumann BC on the brane naturally leads to the vanishing perpendicular velocity $u^w \, \hat{=} \, 0$ and no other conditions on the other components.
This is the usual no-slip condition for ideal fluid, consistent with the phenomenological argument \cite{Landau:FM}.
For viscous fluid, the Neumann BC on the brane leads to additional conditions on the parallel velocity $u^i$ and the inverse temperature $b$, which are given in the form of Neumann boundary condition: $\pa_w u^i \, \hat{=} \, 0$ and $\pa_w b \, \hat{=} \, 0$.
This is not usually found in the phenomenological argument of fluid mechanics and can be regarded as our new results.
The Dirichlet BC on the brane leads to Dirichlet BC for all spacial components of the velocity and the inverse temperature, even in the case of ideal fluid.
This result seems weird from the phenomenological argument of fluid mechanics, which states that for ideal fluid without viscosity, the parallel velocity $u^i$ cannot have any boundary condition.
For the Conformal BC, we have investigated the case where the perturbation is induced by diffeomorphism on the brane.
However, in this case, we could not find any consistent boundary conditions on the fluid velocity.
It seems necessary to find another way to induce perturbation in this case.
Some possibilities are as follows.
(i) We considered the simplest brane embedding solution $w(r)=$const, but we can try to study other non-trivial brane embedding solutions.
(ii) We can also study the perturbation of the metric induced by the fluctuation of the brane embedding.
(iii) We did not introduce intrinsic coordinates on the brane, but as we discussed in Appendix~\ref{app:non-trivial-CBC} we can also study a nont-rivial conformal factor that depends on intrinsic coordinates on the brane.

For non-zero brene tension $T \ne 0$, we only studied ideal fluid with the Neumann boundary condition.
For $d>2$, there is no unique brane location found from the Neumann boundary condition, so we focused on the $d=2$ case.
For ideal fluid in $d=2$, the boundary condition on the brane naturally leads to the vanishing perpendicular velocity $u^w \, \hat{=} \, 0$ as in the tensionless case.

For a future direction, it is interesting to apply our method to charged fluid.
In the context of the fluid/gravity correspondence, charged fluid was studied in \cite{Erdmenger:2008rm, Banerjee:2008th}, and in the AdS/BCFT correspondence gauge fields were studied in \cite{Suzuki:2024cqy}.
One interesting aspect of charged fluid in our program of deriving fluid boundary condition from AdS/BCFT is that even though the gauge field boundary condition on the end-of-the-world brane seems to be independent from the boundary condition of metric in the context of the AdS/BCFT, the fluid/gravity correspondence seems to relate these two boundary conditions due to the appearance of the fluid velocity in the corresponding conserved current.

Another possible direction would be to study the case with brane-localized field as discussed in \cite{Kanda:2023zse}.
It would also be interesting to connect our finite-boundary perspective in fluid/gravity and CBCs with the recent “stretched horizon limit” viewpoint \cite{Anninos:2025zgr}.
As we mentioned in section~\ref{sec:brane-tension}, this might allow us to construct a consistent brane profile with non-zero brane tension in higher dimension.

\section*{Acknowledgements}

We are grateful to Takaaki Ishii, Keisuke Izumi, Taishi Kawamoto, Tetsuya Shiromizu, Tadashi Takayanagi, Norihiro Tanahashi and Tomonori Ugajin for useful discussion.
The work of KS is supported by JSPS KAKENHI Grant No.~23K13105.

\appendix
\section{CBCs with brane intrinsic coordinates}
\label{app:non-trivial-CBC}
In this appendix, we repeat the discussion in section 2.1 of \cite{Galante:2025tnt}, but with a different direction of the boundary.
We study the same bulk $d+1$-dimensional metric
	\begin{align}
		ds_{d+1}^2 \, = \, - r^2 f(br) dt^2 \, + \, \frac{dr^2}{r^2 f(br)} + r^2 dx^i dx_i \, , 
	\label{eq:bulk-metric}
	\end{align}
where $i=1, \cdots, d-1$. In particular we denote $x^{d-1} = w$. In this $d+1$-dimensional bulk spacetime, we put a co-dimension 1 end-of-the-world brane $Q$, and we denote the brane location by
	\begin{align}
		w \, = \, w_*(r) \, .
	\label{eq:w_*}
	\end{align}
This means that the brane $Q$ penetrates the black hole horizon.
Therefore, it is natural to impose the boundary condition on the brane as a $d$-dimensional black hole metric, up to a conformal factor
	\begin{align}
		ds_Q^2 \ \hat{=} \ e^{2\a(\r)} \left( - \r^2 f(b \r e^{\a(\r)}) dt^2 \, + \, \frac{d\r^2}{\r^2 f(b \r e^{\a(\r)})} + \r^2 dx^a dx_a \right) \, , 
	\label{eq:brane-metric}
	\end{align}
where $\r$ is the radial coordinate on the brane and $a=1, \cdots, d-2$.
From the bulk metric (\ref{eq:bulk-metric}) and using the brane profile (\ref{eq:w_*}), we find the induced metric on the brane as
	\begin{align}
		ds_Q^2 \, = \, - r^2 f(br) dt^2 \, + \, \left( \frac{1}{r^2 f(br)} \, + \, r^2 \r'^2 (\partial_{\r} w_*)^2 \right) dr^2 + r^2 dx^a dx_a \, , 
	\end{align}
where the prime denotes a derivative with respect to $r$.
By comparing the $(t,t)$ and $(a, a)$ components of (\ref{eq:brane-metric}), we find 
	\begin{align}
		r \, = \, \r e^{\a(\r)} \, .
	\end{align}
Then, the remaining $(r,r)$ component gives a relationship between the brane profile $w_*(\r)$ and the conformal factor $\a(\r)$:
	\begin{align}
		\pa_\r w_* \ \hat{=} \ \frac{e^{-\a}}{\r^2} \sqrt{\frac{e^{2\a} - (1+\r \pa_\r \a)^2}{f(b \r e^\a)}} \, .
	\end{align}
Once we find the conformal factor $\a(\r)$, we can determine the brane profile $w_*(\r)$ from this equation. 

Next, the conformal factor $\a(\r)$ is fixed by the boundary condition $K \, \hat{=} \, 0$.
The unit normal vector on the brane is given by
	\begin{align}
		n_\m \, = \, \mathcal{N} \big\{ 0, \, - w_*', \, \vec{0}, \, 1 \big\} \, , \qquad \mathcal{N} \, = \, \frac{r}{\sqrt{1+r^4 f w_*'^2}} \, .
	\end{align}
Since the projector is given by
	\begin{align}
		e_{i}^{\m} \, = \, \frac{\pa x^\m}{\pa x^i} \, , \qquad e_{t}^{\m} \, = \,  \big\{ 1, \, 0, \, \vec{0}, \, 0 \big\} \, , \qquad
		e_{\r}^{\m} \, = \,  \big\{ 0, \, \pa_\r r, \, \vec{0}, \, \pa_\r w \big\} \, , \qquad e_{a}^{b} \, = \, \d_{a}^{b} \, ,
	\end{align}
in particular we have 
	\begin{align}
		K_{\r\r} \, = \, (\pa_\r r)^2 K_{rr} \, + \, (\pa_\r r)(\pa_\r w) (K_{rw} + K_{wr}) \, + \, (\pa_\r w)^2 K_{ww} \, .
	\end{align}
The extrinsic curvature is explicitly given by
	\begin{align}
		K_{tt} \, &= \, - \frac{\mathcal{N} F F'}{2} \, w_*' \, , \\
		K_{rr} \, &= \, - \frac{\mathcal{N}^3}{2r^3} \Big( r^3 F' w_*'^2 - 2w_*' - 2r w_*'' - r^3 F w_*' w_*'' \Big) \, - \, \frac{\mathcal{N} F' w_*'}{2F} \, , \\
		K_{rw} \, &= \, - \frac{\mathcal{N}^3}{2r^3} \Big( 2 - r^3 F' w_*' + r^3 F w_*'' \Big) \, - \, \frac{\mathcal{N}}{r} \, , \\
		K_{wr} \, &= \, - \frac{\mathcal{N}}{r} \, , \\
		K_{ww} \, &= \, - r F \mathcal{N} w_*' \, , \\
		K_{xx} \, &= \, - r F \mathcal{N} w_*' \, ,
	\end{align}
where $F(r)=r^2 f(br)$.
From these, we find 
	\begin{align}
		K_{\r\r} \, = \, - \mathcal{N} (\pa_\r r)^2 w_*' \left[ \frac{\mathcal{N}^2 w_*''}{r^2 w_*'} \, + \, \frac{F'}{2F} \, + \, \frac{2}{r} \, + \, r F w_*'^2 \right] \, .
	\end{align}
In order to write the equation in terms of $\pa_\r \a$ and $\pa_\r^2 \a$, let us introduce 
	\begin{align}
		\D \, := \, 1 + \r \pa_\r \a \, ,
	\end{align}
such that
	\begin{align}
		\frac{dr}{d\r} \, = \, \D \, e^\a \, , \qquad \mathcal{N} \, = \, \r \, \D \, , \qquad \frac{\pa w_*}{\pa \r} \, = \, \frac{e^{-\a}}{\r^2} \sqrt{\frac{e^{2\a} - \D^2}{f}} \, .
	\end{align}
From these, we can express 
	\begin{align}
		\frac{w_*''}{w_*'} \, = \, \frac{e^{-\a}}{\D} \left[ \frac{e^{2\a} \pa_\r \a - \D \pa_\r \D}{e^{2\a} - \D^2} \, - \, \frac{e^{\a} \D f'}{2f} \, - \, \frac{2\D}{\r} \, - \, \frac{\pa_\r \D}{\D} \right] \, ,
	\end{align}
and 
	\begin{align}
		K_{\r\r} \, = \, - \mathcal{N} (\pa_\r r)^2 w_*' \left[ \frac{\D}{e^{3\a}} \left( \frac{e^{2\a} \pa_\r \a - \D \pa_\r \D}{e^{2\a} - \D^2} \, - \, \frac{e^\a \D f'}{2f} \, - \, \frac{2\D}{\r} \, - \, \frac{\pa_\r \D}{\D} \right)
		\, + \, \frac{f'}{2f} \, + \, \frac{e^{2\a} + 2\D^2}{\r e^\a \D} \right] \, .
	\end{align}
Finally the trace of the extrinsic curvature is
	\begin{align}
		K \, &= \, h^{tt} K_{tt} \, + \, h^{\r\r} K_{\r\r} \, + \, (d-2) h^{xx} K_{xx} \nn\\[4pt]
		&= \, \mathcal{N} f w_*' \bigg[ \frac{\r^2\D^3}{e^{3\a}} \left( \frac{\D \pa_\r \D - e^{2\a} \pa_\r \a}{e^{2\a} - \D^2} \, + \, \frac{e^\a \D f'}{2f} \, + \, \frac{e^{3\a} f'}{2f \D} \, + \, \frac{2\D}{\r} \, + \, \frac{\pa_\r \D}{\D} \, - \, \frac{2e^{2\a}}{\r \D} \right) \nn\\
        &\hspace{220pt} + \, \frac{\r^2 e^{2\a} f'}{2f} \, + \, (d-2) \r e^\a \bigg] \, . 
	\end{align}
Therefore, the differential equation which determines the conformal factor $\a(\r)$ is given by
	\begin{align}
		0 \, &= \, \frac{\r^2\D^3}{e^{3\a}} \left( \frac{\D \pa_\r \D - e^{2\a} \pa_\r \a}{e^{2\a} - \D^2} \, + \, \frac{e^\a \D f'}{2f} \, + \, \frac{e^{3\a} f'}{2f \D} \, + \, \frac{2\D}{\r} \, + \, \frac{\pa_\r \D}{\D} \, - \, \frac{2e^{2\a}}{\r \D} \right) \nn\\
        &\hspace{180pt} + \, \frac{\r^2 e^{2\a} f'}{2f} \, + \, (d-2) \r e^\a \, .
	\end{align}

\section{Power counting and the $\mathcal{O}(\epsilon)$ conformal Killing equation}
\label{app:CKEcounting}

In this appendix we justify the reduction used in Sec.~4.2.1.
We organize the induced metric by separating the zeroth-order part $g_q$ and the part $h_q$ that is
linear in derivative data:
\begin{align}
h(r,u,b,\partial u,\partial b)_{\alpha \beta}
&=g_{q}(r,u(x),b(x))_{\alpha \beta}
+h_{q}(r,\partial u(x),\partial b(x))_{\alpha \beta},
\label{eq:h_split_app}\\
g_{q}(r,u,b)_{ir}&=-u_{i},
\qquad
h_{q}(r,u,b)_{ir}=0,
\label{eq:gq_hq_ir_app}\\
g_{q}(r,u,b)_{ij}
&=r^{2}\left(\eta_{i j}+(1-f(b r)) u_{i} u_{j}\right),
\label{eq:gq_ij_app}\\
h_{q}(r,u,b)_{ij}
&=2 r^{2} b F(b r) \sigma_{i j }
+\frac{2}{d-1} r u_{i} u_{j} \partial_{\lambda} u^{\lambda}
-r\, u^{\lambda} \partial_{\lambda}\!\left(u_{i} u_{j}\right).
\label{eq:hq_ij_app}
\end{align}

We use the counting
\begin{align}
\partial_r=\mathcal{O}(\epsilon^0),\qquad
\partial_i u,\ \partial_i b=\mathcal{O}(\epsilon),\qquad
\xi^r,\ \xi^i=\mathcal{O}(\epsilon),\qquad
\sigma=\mathcal{O}(\epsilon),
\label{eq:counting_app}
\end{align}
so that $g_q=\mathcal{O}(\epsilon^0)$ and $h_q=\mathcal{O}(\epsilon)$.
The conformal Killing equation on the brane,
\begin{align}
\mathcal{L}_{\xi}h_{\alpha \beta}=2\sigma\,h_{\alpha \beta},
\label{eq:CKE_app}
\end{align}
can then be expanded order by order using
\begin{align}
\mathcal{L}_{\xi}h_{\alpha \beta}
=\xi^{ \gamma }\partial_{ \gamma }h _{\alpha \beta}
+h _{\alpha  \gamma }\partial_{\beta}\xi^{ \gamma }
+h _{ \gamma \beta}\partial_{\alpha }\xi^{ \gamma }.
\label{eq:Lie_app}
\end{align}
With the decomposition \eqref{eq:h_split_app}--\eqref{eq:hq_ij_app} and the counting
\eqref{eq:counting_app}, we have $g_q=\mathcal{O}(\epsilon^0)$ and $h_q=\mathcal{O}(\epsilon)$, while
$\xi^\alpha=\mathcal{O}(\epsilon)$ and $\sigma=\mathcal{O}(\epsilon)$.
Substituting \eqref{eq:h_split_app} into the conformal Killing equation \eqref{eq:CKE_app} and using
\eqref{eq:Lie_app}, we can organize the equation order by order.

At $\mathcal{O}(\epsilon)$, only the $g_q$-sector contributes:
\begin{align}
\mathcal{O}(\epsilon):\quad
\xi^{r}\partial_{r}{g_{q} }_{\alpha \beta}
+{g_{q}}_{\alpha  \gamma }\partial_{\beta}\xi^{ \gamma }
+{g_{q}}_{ \gamma \beta}\partial_{\alpha }\xi^{ \gamma }
=2\sigma\, {g_{q}}_{\alpha \beta}.
\label{eq:CKE_Oe_app}
\end{align}

At the next order, $\mathcal{O}(\epsilon^2)$, the derivative data and the $\mathcal{O}(\epsilon)$
correction $h_q$ start to contribute, and we obtain
\begin{align}
\mathcal{O}(\epsilon^2):\quad
&\xi^{k}\partial_{k}{g_{q} }_{\alpha \beta}
+\xi^{r}\partial_{r}{h_{q} }_{\alpha \beta}
+{h_{q} }_{\alpha  \gamma }\partial_{\beta}\xi^{ \gamma }
+{h_{q} }_{ \gamma \beta}\partial_{\alpha }\xi^{ \gamma }
+\cdots
\notag\\
&\qquad\qquad
=2\sigma\, {h_{q}}_{\alpha \beta}+2\sigma^2\, {g_{q} }_{\alpha \beta}.
\label{eq:CKE_Oe2_app}
\end{align}
Here the term $\xi^{k}\partial_{k}g_q$ is $\mathcal{O}(\epsilon^2)$ because the brane-coordinate
dependence of $g_q$ arises only through $u(x)$ and $b(x)$, so $\partial_k g_q\sim
(\partial_k u,\partial_k b)=\mathcal{O}(\epsilon)$ under \eqref{eq:counting_app}, and multiplying by
$\xi^k=\mathcal{O}(\epsilon)$ gives $\xi^k\partial_k g_q=\mathcal{O}(\epsilon^2)$. Likewise,
$\xi^r\partial_r h_q$ and $h_q\,\partial\xi$ are $\mathcal{O}(\epsilon^2)$ since
$h_q=\mathcal{O}(\epsilon)$ and $\partial\xi=\mathcal{O}(\epsilon)$, while the right-hand side contains
both $2\sigma h_q=\mathcal{O}(\epsilon^2)$ and the quadratic Weyl term $2\sigma^2 g_q=\mathcal{O}(\epsilon^2)$.

Therefore, the leading constraint analyzed in the main text is Eq.~\eqref{eq:CKE_Oe_app}, while the
$\mathcal{O}(\epsilon^2)$ equation \eqref{eq:CKE_Oe2_app} governs the first appearance of derivative
(viscous) data.

\section{Fluctuations induced by the fluid parameters in the CBC analysis}
\label{app:second-order-CBC}

In this appendix, we analyze perturbations arising from fluctuations of fluid parameters, namely the fluid velocity $u^\mu$ and inverse temperature $b$, specifically in the context of the conformal boundary condition (CBC). As discussed in section \ref{sec:conformal}, CBC restricts the induced metric fluctuations to be conformally related to the original metric. Here, we investigate explicitly whether the perturbations induced by these fluid parameter fluctuations can satisfy the CBC.

The perturbation in the fluid/gravity correspondence is obtained by introducing fluctuations in the fluid velocity field $u^{\mu}$ and the inverse temperature $b$. In this case, the fluctuations of $u$ and $b$ are obtained by a Taylor expansion that takes into account their coordinate dependence:
\begin{align}
\delta u_{i} \, = \, \Delta x^{k}\partial_{k}u_{i} \, , \qquad
\delta b \, = \, \Delta x^{k}\partial_{k} b \, , 
\end{align}
where we only consider fluctuations on the EOW brane, so that there is no $w$ components.
Thus, note that the order of these fluctuations is the same as in the fluid/gravity correspondence. Here, we consider up to first order $O(\partial)$. First, for the $(i,r)$ component, we have
\begin{align}
-u_{i}\rightarrow -u_{i}-\delta u_{i},
\end{align}
from which it follows that $\delta u_{i}=2\sigma u_{i}$. Next, let us consider the $(i,j)$ component.
\begin{align}
r^{2}\left(\eta_{ij} + \frac{u_{i}u_{j}}{b^{d}r^{d}}   \right)  &\rightarrow r^{2}\left(\eta_{ij} + \frac{u_{i}u_{j}}{b^{d}r^{d}}   \right) \notag\\
&\quad +\, r^{2}\frac{1}{b^{d}r^{d}}\left( \frac{-d\,\delta b}{b}\, u_{i}u_{j}+\delta u_{i}u_{j} +u_{i}\delta u_{j} \right)\\
&=r^{2}\left(\eta_{ij} + \frac{u_{i}u_{j}}{b^{d}r^{d}}   \right) \notag\\
&\quad +\, r^{2}\frac{u_{i}u_{j}}{b^{d}r^{d}}\left( \frac{-d\,\delta b}{b} +4\sigma \right).
\end{align}
This leads to the condition
$$
r^{2}\frac{u_{i}u_{j}}{b^{d}r^{d}}\left( \frac{-d\,\delta b}{b} +4\sigma \right)= 2\sigma\, r^{2}\left(\eta_{ij} + \frac{u_{i}u_{j}}{b^{d}r^{d}}   \right).
$$
Comparing the terms of order $O(r^{-(d-2)})$, we obtain
$$
\delta b= \frac{2}{d}\sigma\, b.
$$
However, since there is no corresponding term of order $O(r^{2})$ on the left-hand side, we require
$$
\sigma\, r^{2}=0.
$$
Because it is impossible to satisfy this for $\sigma \neq 0$, this perturbation cannot satisfy the conformal boundary condition.

\section{Extrinsic curvature for nonzero brane tension case} 
\label{app:extrinsic-curvature}
In this appendix, we summarize explicit results of the extrinsic curvature for the background (\ref{eq:fluidgravityOrder1d=3}) evaluated on the brane configuration (\ref{eq:BTZw}).
{\tiny
\begin{align}
K^{(0)}_{rr}=& \frac{2 T^3\left(1+2 u_v{ }^2\right) u_w{ }^4}{r^2\left(1+2 T \sqrt{1-f T^2} u_w+\left(1-f T^2\right) u_w{ }^2\right)^{5 / 2}}\notag\\
&+\frac{T^2 \sqrt{1-f T^2} u_w\left(-2 u_v{ }^4 u_w{ }^2+2 f  T^2 u_v{ }^2\left(-u_v{ }^2+u_w{ }^4\right)-T^2\left(2+u_v{ }^4+u_w{ }^4\right)\right)\left(2 f +r b f^\prime\right)}{2 r^2\left(1+2 T \sqrt{1-fT^2} u_w+\left(1-f T^2\right) u_w{ }^2\right)^{5 / 2}}\notag\\
&+\frac{T^2\left(-f  T^3\left(u_v^4+u_w^4\right)+2 T\left(-1+f  T^2\right) u_v^2\left(1+(1+f ) u_v^2 u_w^2+u_w^4\right)\right)\left(2 f  +r b f^\prime\right)}{2 r^2\left(1+2 T \sqrt{1-f  T^2} u_w+\left(1-f   T^2\right) u_w^2\right)^{5 / 2}}\notag\\
&-\frac{T^2 \sqrt{1-f T^2} u_v^2 u_w\left(1-f  T^2-2(1+f  ) T^2 u_w^2+2 u_v^2\left(-f T^2+\left(-1+f T^2\right) u_w^2\right)-r T^2 b f^\prime \right)}{r^2\left(1+2 T \sqrt{1-f T^2} u_w+\left(1-f T^2\right) u_w^2\right)^{5 / 2}}\notag\\
&-\frac{T^3\left(2 f  u_v^2\left(-1+f T^2+4 T^2 u_w^4\right)-T^2 u_w^2\left(4 f  +r b f^\prime \right)+\left(-1+f  T^2\right) u_v^4\left(4 fu_w^2+r b f^\prime \right)\right)}{2 r^2\left(1+2 T \sqrt{1-f  T^2} u_w+\left(1-f  T^2\right) u_w^2\right)^{5 / 2}} .
\end{align}
\begin{align}
K^{(0)}_{r v}=&-\frac{\sqrt{1-f  T^2} u_v u_w\left(2\left(1+f ^2 T^2\right) u_v^2+T^2\left(2 f +r(2+f) b f^\prime +(1+f ) u_w^2\left(-2+r b f^\prime \right)\right)\right)}{2\left(1+2 T \sqrt{1-f T^2} u_w+\left(1-f T^2\right) u_w^2\right)^{3 / 2}}\notag\\
&+\frac{T u_v\left(-2 f +\left(2+f \left(2-6 T^2\right)\right) u_w^4-r u_v^4 b f^\prime +u_w^2\left(-2+f u_v^2\left(-4+2(2+f ) T^2+r T^2 b f^\prime  \right)\right)\right)}{2\left(1+2 T \sqrt{1-f T^2} u_w+\left(1-f T^2\right) u_w^2\right)^{3 / 2}}
\end{align}
\begin{align}
K^{(0)}_{rw}=&\frac{\left(\sqrt{1-fT^2} u_v{ }^2+T u_w\right)\left(T \sqrt{1-f T^2} u_w\left(-8 f^2 T^2+2\left(-3+f r g(2+f) T^2\right) u_v^2+2\left(1-fT^2\right)\left(2-f u_v{ }^4+u_v{ }^2 u_w^2\right)\right)\right)}{2\left(1+2 T \sqrt{1-f T^2} u_w+\left(1-f T^2\right) u_w^2\right)^{5 / 2}}\notag\\
&-\frac{r T \sqrt{1-f T^2}\left((1+4 f) T^2-f  T^2 u_v^2+\left(1-f T^2\right) u_v^4\right) u_w\left(\sqrt{1-f T^2} u_v^2+T u_w\right) b f^\prime }{2\left(1+2 T \sqrt{1-f T^2} u_w+\left(1-f T^2\right) u_w^2\right)^{5 / 2}}\notag\\
&+\frac{\left(\sqrt{1-f  T^2} u_v^2+T u_w\right)\left(fT^2\left(2 f ^2 T^2 u_w^2+2\left(1-f T^2\right)\left(2-3 u_w^2\right)\right)-u_v^2\left(2 fT^2+2\left(1-f T^2\right)\left(2 f T^2+\left(1-2 f T^2+f ^2 T^2\right) u_w^2\right)\right)\right)}{2\left(1+2 T \sqrt{1-f  T^2} u_w+\left(1-f T^2\right) u_w^2\right)^{5 / 2}}\notag\\
&+\frac{r T^2\left(\sqrt{1-f T^2} u_v{ }^2+T u_w\right)\left(f  T^2\left(-1+fu_w^2\right)+\left(-1+f T^2\right)\left(3 u_w^2+u_v^2\left(1+f u_w^2\right)\right)\right) b f^\prime }{2\left(1+2 T \sqrt{1-f  T^2} u_w+\left(1-f T^2\right) u_w^2\right)^{5 / 2}}
\end{align}
\begin{align}
K^{(0)}_{vv}=&-\frac{r^2\left(fT+\sqrt{1-f T^2} u_w\right)\left(-2 u_w^2+u_v^2\left(2 f +r b f^\prime \right)\right)}{2 \sqrt{1+2 T \sqrt{1-fT^2}} u_w+\left(1-f  T^2\right) u_w^2}
\end{align}
\begin{align}
K^{(0)}_{ww}=&-\frac{r^2T^2 \sqrt{1-f  T^2} u_w\left(\left(-2+3 f^2 T^2\right) u_w^2-2 f T^2\left(f-u_w^2\right)\right)\left(2 f+r b f^\prime \right)}{2\left(1+2 T \sqrt{1-fT^2} u_w+\left(1-f  T^2\right) u_w^2\right)^{5 / 2}}\notag \\
&-\frac{r^2 \left(1-fT^2\right)^{3/2}  u_w\left(\left(1-f T^2\right) u_v^4 u_w^2+T^2u_w^4+T^2u_v^2\left(-1+(1+2 f) u_w^2\right)\right)\left(2 f+r b f^\prime \right)}{2\left(1+2 T \sqrt{1-fT^2} u_w+\left(1-f  T^2\right) u_w^2\right)^{5 / 2}}\notag \\
&+\frac{r^2 T\left(-f^3 T^4 u_w^2-\left(1-fT^2\right)^2 u_v^2 u_w^2\left(f   u_v^2+2 u_w^2\right)+fT^2\left(1-f  T^2\right)\left(u_w^2+\left(1-u_w^2\right)\left(u_v^2+3 u_w^2\right)\right)\right)\left(2 f  +r b f^\prime \right)}{2\left(1+2 T \sqrt{1-f T^2} u_w+\left(1-f T^2\right) u_w^2\right)^{5 / 2}}\notag\\
&+\frac{r^2 T \sqrt{1-fT^2} u_w\left(f T+\sqrt{1-f T^2} u_w\right)\left(-4+4 f T^2-4 u_v^2\left(-2+f  T^2+\left(-1+f T^2\right) u_w^2\right)+r T^2 b f^\prime \right)}{2\left(1+2 T \sqrt{1-f T^2} u_w+\left(1-f  T^2\right) u_w^2\right)^{5 / 2}}\notag\\
&+\frac{r^2\left(f T+\sqrt{1-f T^2} u_w\right)\left(2 u_v^2\left(f T^2\left(2-f T^2\right)+2\left(1+(1-2 f) T^2+(-1+f) f T^4\right) u_w^2+\left(-1+f T^2\right)^2 u_w^4\right)\right)}{2\left(1+2 T \sqrt{1-f  T^2} u_w+\left(1-fT^2\right) u_w^2\right)^{5 / 2}}\notag\\
&+\frac{r^2f T^2\left(f T+\sqrt{1-f T^2} u_w\right)\left(-4+4 f T^2+r T^2 b f^\prime\right)}{2\left(1+2 T \sqrt{1-f  T^2} u_w+\left(1-fT^2\right) u_w^2\right)^{5 / 2}}
\end{align}
\begin{align}
K^{(0)}_{vw}=&0
\end{align}
\begin{align}
K^{(0)}= &\frac{T^2 \sqrt{1-f  T^2} u_w\left(f -2 u_w^2+2(-1+f) u_w^4\right)}{\left(\left(1-f T^2\right) u_v^2+T\left(f T+2 \sqrt{1-f T^2} u_w\right)\right)^{3 / 2}}-\frac{T^2\left(-1+f T^2\right) u_v^2 u_w\left(\sqrt{1-f T^2}+(T-f T) u_w\right)\left(f  u_v^2-2 u_w^2\right)\left(2 f +r b f^\prime \right)}{\left(\left(1-f  T^2\right) u_v^2+T\left(f T+2 \sqrt{1-f  T^2} u_w\right)\right)^{5 / 2}}\notag\\
&+\frac{T^2 \sqrt{1-f T^2} u_w\left(5 f T^2+8 u_v^2 u_w^2+2 f ^2 T^2 u_v^2\left(1-4 u_w^2\right)+\left(1-f T^2\right) u_v^2\left(7+2 u_v^2\left(f +2(1-f r g) u_w^2\right)\right)\right)\left(2 f +r b f^\prime \right)}{2\left(\left(1-f  T^2\right) u_v^2+T\left(f T+2 \sqrt{1-f T^2} u_w\right)\right)^{5 / 2}}\notag\\
&+\frac{T\left(f ^2 T^4+2 f T^2\left(4-f (3+f ) T^2\right) u_v^2 u_w^2+\left(1-f T^2\right)\left(3 f T^2 u_v^2+6 T^2 u_w^2+8(1-f) T^2 u_v^2 u_w^4\right)\right)\left(2 f +r b f^\prime \right)}{2\left(\left(1-f T^2\right) u_v^2+T\left(f  T+2 \sqrt{1-f T^2} u_w\right)\right)^{5 / 2}}\notag\\
&+\frac{T\left(f ^2 T^4+2 f T^2\left(4-f (3+f ) T^2\right) u_v^2 u_w^2+2 u_v^4\left(1-f T^2\right)\left(1-fT^2+\left(2-f (1+f ) T^2\right) u_w^2\right)\right)\left(2 f +r b f^\prime \right)}{2\left(\left(1-f T^2\right) u_v^2+T\left(f  T+2 \sqrt{1-f T^2} u_w\right)\right)^{5 / 2}}\notag\\
&+\frac{T\left(2 f^2 T^2-4(1-f) u_v^2 u_w^2-4\left(1-f T^2\right) u_w^2\left(-1+u_w^2\right)+r T^2\left(fu_v{ }^2-u_w^2\right) b f^\prime\right)}{2\left(\left(1-f T^2\right) u_v^2+T\left(f T+2 \sqrt{1-f T^2} u_w\right)\right)^{3 / 2}}\notag\\
&+\frac{\sqrt{1-f T^2} u_v^2 u_w\left(2\left(-1+fT^2\right) u_v^2\left(1-f T^2+(-1+f ) T^2 u_w^2\right)+T^2 u_w^2\left(-2+2 T^2-2(-1+f)^2 T^2+\left(1-f  T^2\right)\left(4 f -r b f^\prime\right)\right)\right)}{\left(\left(1-f T^2\right) u_v^2+T\left(fT+2 \sqrt{1-f  T^2} u_w\right)\right)^{5 / 2}}
\end{align}
}

\bibliographystyle{JHEP}
\bibliography{Refs}

@article{Maldacena:1997re,
    author = "Maldacena, Juan Martin",
    title = "{The Large $N$ limit of superconformal field theories and supergravity}",
    eprint = "hep-th/9711200",
    archivePrefix = "arXiv",
    reportNumber = "HUTP-97-A097, HUTP-98-A097",
    doi = "10.4310/ATMP.1998.v2.n2.a1",
    journal = "Adv. Theor. Math. Phys.",
    volume = "2",
    pages = "231--252",
    year = "1998"
}

@article{Gubser:1998bc,
    author = "Gubser, S. S. and Klebanov, Igor R. and Polyakov, Alexander M.",
    title = "{Gauge theory correlators from noncritical string theory}",
    eprint = "hep-th/9802109",
    archivePrefix = "arXiv",
    reportNumber = "PUPT-1767",
    doi = "10.1016/S0370-2693(98)00377-3",
    journal = "Phys. Lett. B",
    volume = "428",
    pages = "105--114",
    year = "1998"
}

@article{Witten:1998qj,
    author = "Witten, Edward",
    title = "{Anti de Sitter space and holography}",
    eprint = "hep-th/9802150",
    archivePrefix = "arXiv",
    reportNumber = "IASSNS-HEP-98-15",
    doi = "10.4310/ATMP.1998.v2.n2.a2",
    journal = "Adv. Theor. Math. Phys.",
    volume = "2",
    pages = "253--291",
    year = "1998"
}

@article{Policastro:2002se,
    author = "Policastro, Giuseppe and Son, Dam T. and Starinets, Andrei O.",
    title = "{From AdS / CFT correspondence to hydrodynamics}",
    eprint = "hep-th/0205052",
    archivePrefix = "arXiv",
    reportNumber = "INT-PUB-02-32",
    doi = "10.1088/1126-6708/2002/09/043",
    journal = "JHEP",
    volume = "09",
    pages = "043",
    year = "2002"
}

@article{Policastro:2002tn,
    author = "Policastro, Giuseppe and Son, Dam T. and Starinets, Andrei O.",
    title = "{From AdS / CFT correspondence to hydrodynamics. 2. Sound waves}",
    eprint = "hep-th/0210220",
    archivePrefix = "arXiv",
    reportNumber = "DAMTP-2002-129, INT-PUB-02-47",
    doi = "10.1088/1126-6708/2002/12/054",
    journal = "JHEP",
    volume = "12",
    pages = "054",
    year = "2002"
}

@article{Shuryak:2003xe,
    author = "Shuryak, Edward",
    editor = "Faessler, A.",
    title = "{Why does the quark gluon plasma at RHIC behave as a nearly ideal fluid?}",
    eprint = "hep-ph/0312227",
    archivePrefix = "arXiv",
    doi = "10.1016/j.ppnp.2004.02.025",
    journal = "Prog. Part. Nucl. Phys.",
    volume = "53",
    pages = "273--303",
    year = "2004"
}

@article{Shuryak:2004cy,
    author = "Shuryak, Edward V.",
    editor = "Rischke, D. and Levin, G.",
    title = "{What RHIC experiments and theory tell us about properties of quark-gluon plasma?}",
    eprint = "hep-ph/0405066",
    archivePrefix = "arXiv",
    doi = "10.1016/j.nuclphysa.2004.10.022",
    journal = "Nucl. Phys. A",
    volume = "750",
    pages = "64--83",
    year = "2005"
}

@inproceedings{Shuryak:2006se,
author = "Shuryak, E. V.",
title = "{Strongly Coupled Quark-Gluon Plasma: The Status Report}",
booktitle = "{7th Workshop on Continuous Advances in QCD}",
eprint = "hep-ph/0608177",
archivePrefix = "arXiv",
doi = "10.1142/9789812708267_0001",
pages = "3--16",
month = "8",
year = "2006"
}

@article{Kovtun:2004de,
author = "Kovtun, P. and Son, Dan T. and Starinets, Andrei O.",
title = "{Viscosity in strongly interacting quantum field theories from black hole physics}",
eprint = "hep-th/0405231",
archivePrefix = "arXiv",
reportNumber = "INT-PUB-04-09, UW-PT-04-04",
doi = "10.1103/PhysRevLett.94.111601",
journal = "Phys. Rev. Lett.",
volume = "94",
pages = "111601",
year = "2005"
}

@article{Bhattacharyya:2007vjd,
    author = "Bhattacharyya, Sayantani and Hubeny, Veronika E and Minwalla, Shiraz and Rangamani, Mukund",
    title = "{Nonlinear Fluid Dynamics from Gravity}",
    eprint = "0712.2456",
    archivePrefix = "arXiv",
    primaryClass = "hep-th",
    reportNumber = "TIFR-TH-07-44, DCPT-07-73, NI07097",
    doi = "10.1088/1126-6708/2008/02/045",
    journal = "JHEP",
    volume = "02",
    pages = "045",
    year = "2008"
}

@article{Bhattacharyya:2008mz,
    author = "Bhattacharyya, Sayantani and Loganayagam, R. and Mandal, Ipsita and Minwalla, Shiraz and Sharma, Ankit",
    title = "{Conformal Nonlinear Fluid Dynamics from Gravity in Arbitrary Dimensions}",
    eprint = "0809.4272",
    archivePrefix = "arXiv",
    primaryClass = "hep-th",
    reportNumber = "TIFR-TH-08-38",
    doi = "10.1088/1126-6708/2008/12/116",
    journal = "JHEP",
    volume = "12",
    pages = "116",
    year = "2008"
}

@inproceedings{Ambrosetti:2008mt,
    author = "Ambrosetti, Nicola and Charbonneau, James and Weinfurtner, Silke",
    title = "{The Fluid/gravity correspondence: Lectures notes from the 2008 Summer School on Particles, Fields, and Strings}",
    booktitle = "{6th Summer School on Particles, Fields and Strings}",
    eprint = "0810.2631",
    archivePrefix = "arXiv",
    primaryClass = "gr-qc",
    month = "10",
    year = "2008"
}

@article{Haack:2008cp,
    author = "Haack, Michael and Yarom, Amos",
    title = "{Nonlinear viscous hydrodynamics in various dimensions using AdS/CFT}",
    eprint = "0806.4602",
    archivePrefix = "arXiv",
    primaryClass = "hep-th",
    reportNumber = "LMU-ASC-39-08",
    doi = "10.1088/1126-6708/2008/10/063",
    journal = "JHEP",
    volume = "10",
    pages = "063",
    year = "2008"
}

@inproceedings{Hubeny:2011hd,
    author = "Hubeny, Veronika E. and Minwalla, Shiraz and Rangamani, Mukund",
    title = "{The fluid/gravity correspondence}",
    booktitle = "{Theoretical Advanced Study Institute in Elementary Particle Physics}: {String theory and its Applications: From meV to the Planck Scale}",
    eprint = "1107.5780",
    archivePrefix = "arXiv",
    primaryClass = "hep-th",
    pages = "348--383",
    year = "2012"
}

@article{Witten:2018lgb,
    author = "Witten, Edward",
    title = "{A note on boundary conditions in Euclidean gravity}",
    eprint = "1805.11559",
    archivePrefix = "arXiv",
    primaryClass = "hep-th",
    doi = "10.1142/S0129055X21400043",
    journal = "Rev. Math. Phys.",
    volume = "33",
    number = "10",
    pages = "2140004",
    year = "2021"
}

@book{levlandaufluid,
	title = {Fluid Mechanics: Landau and Lifshitz: Course of Theoretical Physics},
	author = {{Lev  Davidovich Landau} and {Evgeny Mikhailovich Lifshitz}},
}

@article{Karch:2000gx,
    author = "Karch, Andreas and Randall, Lisa",
    title = "{Open and closed string interpretation of SUSY CFT's on branes with boundaries}",
    eprint = "hep-th/0105132",
    archivePrefix = "arXiv",
    reportNumber = "MIT-CTP-3146",
    doi = "10.1088/1126-6708/2001/06/063",
    journal = "JHEP",
    volume = "06",
    pages = "063",
    year = "2001"
}

@article{Takayanagi:2011zk,
    author = "Takayanagi, Tadashi",
    title = "{Holographic Dual of BCFT}",
    eprint = "1105.5165",
    archivePrefix = "arXiv",
    primaryClass = "hep-th",
    reportNumber = "IPMU11-0091",
    doi = "10.1103/PhysRevLett.107.101602",
    journal = "Phys. Rev. Lett.",
    volume = "107",
    pages = "101602",
    year = "2011"
}

@article{Fujita:2011fp,
    author = "Fujita, Mitsutoshi and Takayanagi, Tadashi and Tonni, Erik",
    title = "{Aspects of AdS/BCFT}",
    eprint = "1108.5152",
    archivePrefix = "arXiv",
    primaryClass = "hep-th",
    reportNumber = "IPMU-11-0136, MIT-CTP-4289",
    doi = "10.1007/JHEP11(2011)043",
    journal = "JHEP",
    volume = "11",
    pages = "043",
    year = "2011"
}

@article{Nozaki:2012qd,
    author = "Nozaki, Masahiro and Takayanagi, Tadashi and Ugajin, Tomonori",
    title = "{Central Charges for BCFTs and Holography}",
    eprint = "1205.1573",
    archivePrefix = "arXiv",
    primaryClass = "hep-th",
    reportNumber = "YITP-12-42, IPMU12-0087",
    doi = "10.1007/JHEP06(2012)066",
    journal = "JHEP",
    volume = "06",
    pages = "066",
    year = "2012"
}

@article{Miyaji:2021ktr,
    author = "Miyaji, Masamichi and Takayanagi, Tadashi and Ugajin, Tomonori",
    title = "{Spectrum of End of the World Branes in Holographic BCFTs}",
    eprint = "2103.06893",
    archivePrefix = "arXiv",
    primaryClass = "hep-th",
    reportNumber = "YITP-21-19, IPMU21-0017",
    doi = "10.1007/JHEP06(2021)023",
    journal = "JHEP",
    volume = "06",
    pages = "023",
    year = "2021"
}

@article{Randall:1999ee,
    author = "Randall, Lisa and Sundrum, Raman",
    title = "{A Large mass hierarchy from a small extra dimension}",
    eprint = "hep-ph/9905221",
    archivePrefix = "arXiv",
    reportNumber = "MIT-CTP-2860, PUPT-1860, BUHEP-99-9",
    doi = "10.1103/PhysRevLett.83.3370",
    journal = "Phys. Rev. Lett.",
    volume = "83",
    pages = "3370--3373",
    year = "1999"
}

@article{Randall:1999vf,
    author = "Randall, Lisa and Sundrum, Raman",
    title = "{An Alternative to compactification}",
    eprint = "hep-th/9906064",
    archivePrefix = "arXiv",
    reportNumber = "MIT-CTP-2874, PUPT-1867, BUHEP-99-13",
    doi = "10.1103/PhysRevLett.83.4690",
    journal = "Phys. Rev. Lett.",
    volume = "83",
    pages = "4690--4693",
    year = "1999"
}

@article{Karch:2000ct,
    author = "Karch, Andreas and Randall, Lisa",
    editor = "Duff, Michael J. and Liu, J. T. and Lu, J.",
    title = "{Locally localized gravity}",
    eprint = "hep-th/0011156",
    archivePrefix = "arXiv",
    reportNumber = "MIT-CTP-3099",
    doi = "10.1088/1126-6708/2001/05/008",
    journal = "JHEP",
    volume = "05",
    pages = "008",
    year = "2001"
}

@article{Anninos:2023epi,
author = "Anninos, Dionysios and Galante, Dami\'an A. and Maneerat, Chawakorn",
title = "{Gravitational observatories}",
eprint = "2310.08648",
archivePrefix = "arXiv",
primaryClass = "hep-th",
doi = "10.1007/JHEP12(2023)024",
journal = "JHEP",
volume = "12",
pages = "024",
year = "2023"
}

@article{Banihashemi:2024yye,
author = "Banihashemi, Batoul and Shaghoulian, Edgar and Shashi, Sanjit",
title = "{Flat space gravity at finite cutoff}",
eprint = "2409.07643",
archivePrefix = "arXiv",
primaryClass = "hep-th",
doi = "10.1088/1361-6382/ada2d7",
journal = "Class. Quant. Grav.",
volume = "42",
number = "3",
pages = "035010",
year = "2025"
}

@article{Miao:2017gyt,
    author = "Miao, Rong-Xin and Chu, Chong-Sun and Guo, Wu-Zhong",
    title = "{New proposal for a holographic boundary conformal field theory}",
    eprint = "1701.04275",
    archivePrefix = "arXiv",
    primaryClass = "hep-th",
    reportNumber = "NCTS-TH-1701",
    doi = "10.1103/PhysRevD.96.046005",
    journal = "Phys. Rev. D",
    volume = "96",
    number = "4",
    pages = "046005",
    year = "2017"
}

@article{Chu:2017aab,
    author = "Chu, Chong-Sun and Miao, Rong-Xin and Guo, Wu-Zhong",
    title = "{On New Proposal for Holographic BCFT}",
    eprint = "1701.07202",
    archivePrefix = "arXiv",
    primaryClass = "hep-th",
    reportNumber = "NCTS-TH-1702",
    doi = "10.1007/JHEP04(2017)089",
    journal = "JHEP",
    volume = "04",
    pages = "089",
    year = "2017"
}

@article{Suzuki:2022xwv,
    author = "Suzuki, Kenta and Takayanagi, Tadashi",
    title = "{BCFT and Islands in two dimensions}",
    eprint = "2202.08462",
    archivePrefix = "arXiv",
    primaryClass = "hep-th",
    reportNumber = "YITP-22-14, IPMU22-0002",
    doi = "10.1007/JHEP06(2022)095",
    journal = "JHEP",
    volume = "06",
    pages = "095",
    year = "2022"
}

@article{Izumi:2022opi,
    author = "Izumi, Keisuke and Shiromizu, Tetsuya and Suzuki, Kenta and Takayanagi, Tadashi and Tanahashi, Norihiro",
    title = "{Brane dynamics of holographic BCFTs}",
    eprint = "2205.15500",
    archivePrefix = "arXiv",
    primaryClass = "hep-th",
    reportNumber = "YITP-22-56, IPMU22-0032",
    doi = "10.1007/JHEP10(2022)050",
    journal = "JHEP",
    volume = "10",
    pages = "050",
    year = "2022"
}

@article{Chu:2018ntx,
    author = "Chu, Chong-Sun and Miao, Rong-Xin",
    title = "{Anomalous Transport in Holographic Boundary Conformal Field Theories}",
    eprint = "1804.01648",
    archivePrefix = "arXiv",
    primaryClass = "hep-th",
    reportNumber = "NCTS-TH-1806, NCTS-TH/1806",
    doi = "10.1007/JHEP07(2018)005",
    journal = "JHEP",
    volume = "07",
    pages = "005",
    year = "2018"
}

@article{Miao:2018qkc,
    author = "Miao, Rong-Xin",
    title = "{Holographic BCFT with Dirichlet Boundary Condition}",
    eprint = "1806.10777",
    archivePrefix = "arXiv",
    primaryClass = "hep-th",
    doi = "10.1007/JHEP02(2019)025",
    journal = "JHEP",
    volume = "02",
    pages = "025",
    year = "2019"
}

@article{Chu:2021mvq,
    author = "Chu, Chong-Sun and Miao, Rong-Xin",
    title = "{Conformal boundary condition and massive gravitons in AdS/BCFT}",
    eprint = "2110.03159",
    archivePrefix = "arXiv",
    primaryClass = "hep-th",
    doi = "10.1007/JHEP01(2022)084",
    journal = "JHEP",
    volume = "01",
    pages = "084",
    year = "2022"
}

@article{Kanda:2023zse,
    author = "Kanda, Hiroki and Sato, Masahide and Suzuki, Yu-ki and Takayanagi, Tadashi and Wei, Zixia",
    title = "{AdS/BCFT with brane-localized scalar field}",
    eprint = "2302.03895",
    archivePrefix = "arXiv",
    primaryClass = "hep-th",
    reportNumber = "YITP-23-07",
    doi = "10.1007/JHEP03(2023)105",
    journal = "JHEP",
    volume = "03",
    pages = "105",
    year = "2023"
}

@article{Erdmenger:2008rm,
    author = "Erdmenger, Johanna and Haack, Michael and Kaminski, Matthias and Yarom, Amos",
    title = "{Fluid dynamics of R-charged black holes}",
    eprint = "0809.2488",
    archivePrefix = "arXiv",
    primaryClass = "hep-th",
    reportNumber = "LMU-ASC-48-08, MPP-2008-116",
    doi = "10.1088/1126-6708/2009/01/055",
    journal = "JHEP",
    volume = "01",
    pages = "055",
    year = "2009"
}

@article{Banerjee:2008th,
    author = "Banerjee, Nabamita and Bhattacharya, Jyotirmoy and Bhattacharyya, Sayantani and Dutta, Suvankar and Loganayagam, R. and Surowka, P.",
    title = "{Hydrodynamics from charged black branes}",
    eprint = "0809.2596",
    archivePrefix = "arXiv",
    primaryClass = "hep-th",
    doi = "10.1007/JHEP01(2011)094",
    journal = "JHEP",
    volume = "01",
    pages = "094",
    year = "2011"
}

@article{Suzuki:2024cqy,
    author = "Suzuki, Kenta",
    title = "{Gauge symmetries and conserved currents in AdS/BCFT}",
    eprint = "2403.07325",
    archivePrefix = "arXiv",
    primaryClass = "hep-th",
    reportNumber = "RUP-24-4",
    doi = "10.1007/JHEP06(2024)137",
    journal = "JHEP",
    volume = "06",
    pages = "137",
    year = "2024"
}

@book{Landau:FM,
  author    = "L.D. Landau and E.M. Lifshitz",
  title     = "Fluid Mechanics",
  publisher = "Pergamon Press",
  year      = "1987",
}

@article{Anderson_2008,
title={On boundary value problems for Einstein metrics},
volume={12},
ISSN={1465-3060},
url={http://dx.doi.org/10.2140/gt.2008.12.2009},
DOI={10.2140/gt.2008.12.2009},
number={4},
journal={Geometry \& Topology},
publisher={Mathematical Sciences Publishers},
author={Anderson, Michael T},
year={2008},
month=jul, pages={2009–2045} }

@article{Liu:2024ymn,
author = "Liu, Xiaoyi and Santos, Jorge E. and Wiseman, Toby",
title = "{New Well-Posed boundary conditions for semi-classical Euclidean gravity}",
eprint = "2402.04308",
archivePrefix = "arXiv",
primaryClass = "hep-th",
doi = "10.1007/JHEP06(2024)044",
journal = "JHEP",
volume = "06",
pages = "044",
year = "2024"
}

@article{Fournodavlos:2021eye,
    author = "Fournodavlos, Grigorios and Smulevici, Jacques",
    title = "{The Initial Boundary Value Problem in General Relativity: The Umbilic Case}",
    eprint = "2104.08851",
    archivePrefix = "arXiv",
    primaryClass = "gr-qc",
    doi = "10.1093/imrn/rnab359",
    journal = "Int. Math. Res. Not.",
    volume = "2023",
    number = "5",
    pages = "3790--3807",
    year = "2023"
}

@article{An:2021fcq,
    author = "An, Zhongshan and Anderson, Michael T.",
    title = "{The initial boundary value problem and quasi-local Hamiltonians in General Relativity}",
    eprint = "2103.15673",
    archivePrefix = "arXiv",
    primaryClass = "gr-qc",
    doi = "10.1088/1361-6382/ac0a86",
    month = "3",
    year = "2021"
}

@article{An:2025rlw,
    author = "An, Zhongshan and Anderson, Michael T.",
    title = "{Well-posed geometric boundary data in General Relativity, I: Conformal-mean curvature boundary data}",
    eprint = "2503.12599",
    archivePrefix = "arXiv",
    primaryClass = "math.AP",
    month = "3",
    year = "2025"
}

@article{An:2025gvr,
    author = "An, Zhongshan and Anderson, Michael T.",
    title = "{Well-posed geometric boundary data in General Relativity, II: Dirichlet boundary data}",
    eprint = "2505.07128",
    archivePrefix = "arXiv",
    primaryClass = "math.AP",
    month = "5",
    year = "2025"
}

@article{Liu:2025xij,
author = "Liu, Xiaoyi and Reall, Harvey S. and Santos, Jorge E. and Wiseman, Toby",
title = "{Ill-posedness of the Cauchy problem for linearized gravity in a cavity with conformal boundary conditions}",
eprint = "2505.20410",
archivePrefix = "arXiv",
primaryClass = "gr-qc",
doi = "10.1088/1361-6382/ae1b60",
journal = "Class. Quant. Grav.",
volume = "42",
number = "23",
pages = "235003",
year = "2025"
}

@article{Anninos:2024wpy,
author = "Anninos, Dionysios and Galante, Dami{\'a}n A. and Maneerat, Chawakorn",
title = "{Cosmological observatories}",
eprint = "2402.04305",
archivePrefix = "arXiv",
primaryClass = "hep-th",
doi = "10.1088/1361-6382/ad5824",
journal = "Class. Quant. Grav.",
volume = "41",
number = "16",
pages = "165009",
year = "2024"
}

@article{Anninos:2024xhc,
author = "Anninos, Dionysios and Arias, Ra{\'u}l and Galante, Dami{\'a}n A. and Maneerat, Chawakorn",
title = "{Gravitational observatories in AdS$_{4}$}",
eprint = "2412.16305",
archivePrefix = "arXiv",
primaryClass = "hep-th",
doi = "10.1007/JHEP07(2025)234",
journal = "JHEP",
volume = "07",
pages = "234",
year = "2025"
}

@article{Galante:2025tnt,
author = "Galante, Dami{\'a}n A. and Maneerat, Chawakorn and Svesko, Andrew",
title = "{Conformal boundaries near extremal black holes}",
eprint = "2504.14003",
archivePrefix = "arXiv",
primaryClass = "hep-th",
doi = "10.1088/1361-6382/ae0408",
journal = "Class. Quant. Grav.",
volume = "42",
number = "19",
pages = "195003",
year = "2025"
}

@article{Anninos:2025zgr,
author = "Anninos, Dionysios and Galante, Dami{\'a}n A. and Georgescu, Silvia and Maneerat, Chawakorn and Svesko, Andrew",
title = "{The Stretched Horizon Limit}",
eprint = "2512.16738",
archivePrefix = "arXiv",
primaryClass = "hep-th",
month = "12",
year = "2025"
}

@article{Kovtun:2012rj,
    author = "Kovtun, Pavel",
    title = "{Lectures on hydrodynamic fluctuations in relativistic theories}",
    eprint = "1205.5040",
    archivePrefix = "arXiv",
    primaryClass = "hep-th",
    doi = "10.1088/1751-8113/45/47/473001",
    journal = "J. Phys. A",
    volume = "45",
    pages = "473001",
    year = "2012"
}


\end{document}